\documentclass[a4paper,11pt]{article}

\usepackage{jcappub} 
\usepackage[T1]{fontenc}
\usepackage{blindtext}
\usepackage[utf8]{inputenc}
\usepackage{graphicx}
\usepackage{subcaption}
\usepackage{dcolumn}
\usepackage{bm}
\usepackage{epsfig}
\usepackage{amsfonts}
\usepackage{amsmath}
\usepackage{amssymb}
\usepackage[usenames]{color}
\usepackage[dvipsnames]{xcolor}
\usepackage{float}
\usepackage{caption}
\usepackage{booktabs}
\usepackage{hyperref}
\usepackage{arydshln}
\usepackage[shortlabels]{enumitem}
\usepackage{multirow}
\usepackage{multicol}
\usepackage[normalem]{ulem}

\arxivnumber{2301.12708} 

\title{\boldmath  A thorough investigation of the prospects of eLISA in addressing the Hubble tension: Fisher Forecast, MCMC and Machine Learning}

\author[*,1]{Rahul Shah,\note{Corresponding author}}
\author[*]{Arko Bhaumik,}
\author[*]{Purba Mukherjee}
\author[*,\dagger]{and Supratik Pal}

\affiliation[*]{Physics and Applied Mathematics Unit, Indian Statistical Institute,\\203, B.T. Road, Kolkata 700 108, India}
\affiliation[\dagger]{Technology Innovation Hub on Data Science, Big Data Analytics and Data Curation,\\
Indian Statistical Institute, 203, B.T. Road, Kolkata 700 108, India}

\emailAdd{rahul.shah.13.97@gmail.com}
\emailAdd{arkobhaumik12@gmail.com}
\emailAdd{purba16@gmail.com}
\emailAdd{supratik@isical.ac.in}

\abstract{
We carry out an in-depth analysis of the capability of the upcoming space-based gravitational wave mission eLISA in addressing the Hubble tension, with a primary focus on observations at intermediate redshifts ($3<z<8$). We consider six different parametrizations representing different classes of cosmological models, which we constrain using the latest datasets of cosmic microwave background (CMB), baryon acoustic oscillations (BAO), and type Ia supernovae (SNIa) observations, in order to find out the up-to-date tensions with direct measurement data. Subsequently, these constraints are used as fiducials to construct mock catalogs for eLISA. We then employ Fisher analysis to forecast the future performance of each model in the context of eLISA. We further implement traditional Markov Chain Monte Carlo (MCMC) to estimate the parameters from the simulated catalogs. Finally, we utilize Gaussian Processes (GP), a machine learning algorithm, for reconstructing the Hubble parameter directly from simulated data. Based on our analysis, we present a thorough comparison of the three methods as forecasting tools. Our Fisher analysis confirms that eLISA would constrain the Hubble constant ($H_0$) at the sub-percent level. MCMC/GP results predict reduced tensions for models/fiducials which are currently harder to reconcile with direct measurements of $H_0$, whereas no significant change occurs for models/fiducials at lesser tensions with the latter. This feature warrants further investigation in this direction.
}

\begin{document}

\maketitle 

\flushbottom


\section{\label{sec:introduction}Introduction}
While the standard $\Lambda$CDM (cosmological constant $+$ cold dark matter) model of cosmology has provided an excellent fit to a wide range of cosmological datasets over the last two decades, state-of-the-art observational facilities in the era of precision cosmology have recently started to shed light on its inadequacies. The latter have manifested in the form of discrepancies or \textit{tensions} between the values of one or more parameter(s) of the baseline $\Lambda$CDM model, inferred from different datasets. Among the key tensions observed so far, perhaps the most serious one is the so-called Hubble tension, which has emerged between the value of the Hubble constant ($H_0$) inferred from cosmic microwave background (CMB) data at a redshift of $z\sim1080$ and direct model-independent measurement of $H_0$ from low-redshift $(z\sim0.01-1)$ Cepheid-calibrated type Ia supernovae (SNIa) data. First identified after the \textit{Planck} 2013 data release \cite{novosyadlyj,hazra}, it has steadily grown into a serious problem in the standard cosmological paradigm that can no longer be overlooked. Latest estimates yield $H_0=67.36\pm0.54$ km s\textsuperscript{-1} Mpc\textsuperscript{-1} from CMB observations by \textit{Planck} 2018 assuming baseline $\Lambda$CDM \cite{Pl18VI}, in contrast to the significantly higher value of $H_0=73.30\pm1.04$ km s\textsuperscript{-1} Mpc\textsuperscript{-1} obtained by Riess et al. (R21) of the Supernovae, H0, for the Equation of State of Dark Energy (SH0ES) collaboration \cite{Riess_2022}. This translates to a strong $\sim 5\sigma$ tension between the results of these early and late time probes of $H_0$. The situation is especially problematic since the \textit{Planck} 2018 best fit value is concordant with $H_0=67.4^{+1.1}_{-1.2}$ km s\textsuperscript{-1} Mpc\textsuperscript{-1} inferred from the Dark Energy Survey Year 1 (DES-Y1) clustering and weak lensing data in conjunction with baryon acoustic oscillations (BAO) datasets from spectroscopic surveys \cite{DESY1BAO}. On the other hand, higher values of $H_0$ are also obtained by several other SNIa studies using alternative distance calibrators such as Tip of the Red Giant Branch (TRGB) stars \cite{trgb1,trgb2,trgb3,trgb4} and Mira variable red giants \cite{mira}. Moving beyond SNIa observations, the Megamaser Cosmology Project (MCP) \cite{mcp1} infers $H_0=73.9\pm3.0$ km s\textsuperscript{-1} Mpc\textsuperscript{-1} from very-long-baseline interferometric studies of extragalactic water masers in orbit around supermassive black holes (SMBHs) \cite{mcp2}, whereas the $H_0$ Lenses in COSMOGRAIL's Wellspring (H0LiCOW) collaboration reports $H_0=73.3^{+1.7}_{-1.8}$ km s\textsuperscript{-1} Mpc\textsuperscript{-1} based on a joint analysis of six gravitationally lensed quasars with measured time delays \cite{holicow}. Clearly, a higher value of $H_0$ is preferred by various direct measurement datasets which do not need to assume any specific cosmological model, unlike what is required for the estimation of $H_0$ from CMB data. Apart from the dominant $H_0$ tension, recent data also reflects the $S_8$ tension (at $\sim2-3\sigma$) between locally measured values of the root mean square density fluctuations of matter \cite{S81,S82,S83,S84,S85} and a higher value estimated from CMB \cite{Pl18VI,Pl18VIII}. Besides, the CMB temperature power spectrum exhibits an anomalous smoothing effect of gravitational lensing quantified in terms of the parameter $A_L$, whose inferred value is at $\sim2.8\sigma$ tension with the $\Lambda$CDM-predicted value of unity \cite{Pl18VII,CMBLensing1,CMBLensing2,CMBLensing3}. Such discrepancies naturally lead the community to explore the prospects of beyond-$\Lambda$CDM models, either within the 6-parameter framework or by introducing one or more parameter(s) on top of the vanilla 6-parameter description.

A plethora of alternative cosmological models has been suggested thus far, with varying degrees of success in addressing the $H_0$ tension based on presently available datasets (for a comprehensive review of examples and their current status see \cite{DiValentino_2021,hubble_buyers_guide,hubble_hunters_guide,To_or_not_to_H0,H0_olympics,Dainotti_2021,Dainotti_2022}). These include diverse proposals to modify both early time and late time cosmological dynamics by introducing additional physics, while at the same time ensuring that the well-established merits of the concordance model are not jeopardized. However, none of them could completely resolve the $H_0$ tension satisfactorily without transferring the tension to another cosmological parameter, or without invoking some very unusual theoretical propositions, or without considering some not-so-convincing data. This is particularly because of the degeneracies between $H_0$ and other cosmological parameters \cite{deg1,deg2,deg3,deg4} as well as a positive correlation between $H_0$ and $\sigma_8$ (see, for example, \cite{Bhattacharyya_2019}) so far as current observational datasets are concerned.

While the chances of resolving the issue completely by an as-yet-unknown cosmological model that can take care of all the virtues of the baseline $\Lambda$CDM model while overcoming its vices (thereby serving as the {\it holy grail} of modern cosmology) is still there, the increasing discontent in the community about the existing datasets in addressing the Hubble tension issue leads us to look beyond current observations. The major hindrance in this direction is the scarcity of direct data at intermediate redshifts beyond $z\sim2$ till date that continues to impede progress when it comes to more precise analysis of these models.  Upcoming missions like the Square Kilometre Array (SKA) \cite{ska} and the Thirty Meter Telescope (TMT) \cite{tmt} aim to probe these redshifts and test the predictions of these models at both background and perturbative levels to much higher precision. However, they are  restricted solely to the electromagnetic (EM) spectrum, having their own observational limitations. Moreover, as already pointed out, current degeneracies between $H_0$ and other cosmological parameters \cite{deg1,deg2,deg3,deg4} complicate the situation and give rise to the necessity of probing these intermediate redshifts via alternative channels in addition to the standard EM observations.

That is precisely where future gravitational wave (GW) detectors come into the picture. Next generation GW observatories, like the ground-based Einstein Telescope (ET) \cite{ET1,ET2,Cao_2022} or the space-based Laser Interferometer Space Antenna (LISA) \cite{eLISA,eLISA1,eLISA2,eLISA3,eLISA4} and the Deci-Hertz Interferometer Gravitational Wave Observatory (DECIGO) \cite{DECIGO1,DECIGO2,DECIGO3}, are expected to play a key role in cosmography in the coming decades. In addition to opening up a new window into the Universe beyond conventional EM astronomy, they will help us probe intermediate redshifts via frequent detections of compact binary coalescence events in the hitherto inaccessible $z\gtrsim2$ range. For our present study, we focus on the capabilities of the currently planned ``evolved LISA'' (eLISA) mission only. Some of the design configurations currently proposed for eLISA should enable it to detect massive black hole binary (MBHB) mergers having electromagnetic counterparts up to $z\sim8$ at the rate of a few events per year, with signal-to-noise ratio (SNR) $>8$ and sky location error below $10$ deg\textsuperscript{2} \cite{eLISA3}. The GW waveform contains enough information to enable direct measurement of the GW luminosity distance to the source \cite{Schutz1986,Holz_2005,Cutler_et_al}, while parallel detections of electromagnetic counterparts to the GW event by other missions would help determine the redshift. The feasibility of such multi-messenger observations has already been demonstrated through the gravitational detection of the binary neutron star merger event GW170817 by the LIGO-VIRGO collaboration \cite{bns}, alongside electromagnetic detection of its gamma-ray burst counterpart GRB 170817A by a number of independent groups of EM observers \cite{grb1,grb2,grb3}. For eLISA, such simultaneous detections can be achieved with the Large Synoptic Survey Telescope (LSST) \cite{lsst} for a sufficiently bright optical counterpart, or with the SKA together with Extremely Large Telescope (ELT) follow-ups if it falls within the radio band \cite{elt}. Equipped with the redshift versus luminosity distance relation of standard sirens, the expansion history can subsequently be studied and precise constraints can be placed on the parameter space of the cosmological model under consideration. This has been demonstrated in a couple of previous works through the well-accepted Fisher matrix forecast and Markov Chain Monte Carlo (MCMC) analyses based on simulated eLISA standard siren catalogs by a model that modifies the early Universe cosmology and by taking into account interactions among two cosmic species (dark matter and dark energy) \cite{Early_Int_DE}, as well as by a model that modifies the late time cosmological scenario \cite{eLISA3}. Forecast on the prospects of null diagnostics in the light of eLISA has also been carried out to some extent \cite{Baral2021}. However, a thorough and methodical forecast analysis considering different types of cosmological models that show moderate to significant levels of promise in addressing the Hubble tension is yet to see the light of day. 

On the other hand, the accumulation and processing of large volumes of data has become the cornerstone of precision cosmology. This entails the need for faster and more efficient computational tools and data handling algorithms. Besides conventional methods of simulation and data analysis, various machine learning (ML) techniques like Gaussian Processes (GP), Genetic Algorithms (GA), and various deep learning algorithms are increasingly being used in different areas of cosmology (for a small body of diverse examples from recent years see \cite{Cai_2017,Schaefer_2018,He_2019,Zhou_2019,Hassan_2020,Belgacem_2020,Yang_2021,Ca_as_Herrera_2021,Zheng_2021,Wang_2021,Aizpuru_2021,Mukherjee_2021,Kacprzak_2022,Bengaly_2022,Alhassan_2023,ML_Hartman,Mukherjee:2022yyq,Sharma_2022_1,Sharma_2022_2}). Gaussian Processes, for example, have already found considerable application in the area of non-parametric reconstructions of various cosmological parameters \cite{Holsclaw_2011,Seikel:2012uu,Shafieloo_2012,Seikel:2013fda}. By the time the next generation cosmological missions go online in the coming decades and start generating enormous amounts of data, the role of sophisticated ML tools, both in their standalone capacity and in conjunction with more conventional data analysis techniques, may prove to be of paramount importance. Thus, it is the need of the hour to thoroughly assess the competence as well as limitations of these ML pipelines when it comes to cosmological data, against those of the more well-established methods widely used by the community. 

In this study, we carry out an in-depth investigation of the prospects of eLISA in addressing the Hubble tension for a few interesting cosmological models in the background. To this end, we consider six different parametrizations representing different classes of models, with 6, (6+1), and (6+2)-parameter descriptions respectively, and find out the up-to-date constraints on the model parameters from the latest cosmological datasets by MCMC analysis, that are found to be lacking in the literature. This helps us compare the models on an equal footing. Then, using these constraints from existing datasets as fiducials, we generate the mock catalogs of eLISA for the investigation that follows. It deserves mention that instead of assuming just the $\Lambda$CDM fiducial to generate the mock catalog (as done in most approaches in the literature), we have considered fiducials motivated from different classes of cosmological models for a more robust and consistent analysis. We then employ a three-pronged methodology by comparing among the results of three distinct approaches, namely: (i) Fisher forecasting, (ii) Markov Chain Monte Carlo (MCMC), and (iii) Gaussian Processes (GP) in machine learning. As is well-known, Fisher and MCMC are standard parametric approaches which depend on the underlying cosmological model, whereas GP is a non-parametric method that does not require the assumption of any particular model. In our scheme of work, we apply the GP regression technique on catalogs constructed by taking the constraints for the different models as fiducials, so that we can examine to what extent the reconstruction process gets affected by different fiducials. We deliberately choose such model-inspired fiducials instead of arbitrary ones for GP in order to lend credence to the reconstruction results from a physical standpoint. This way, our three-pronged methodology helps to achieve a twofold goal. Firstly, it permits a multi-channel analysis of the fiducials inspired from individual cosmological models and their tension-resolving potential based on realistic eLISA mock catalogs. Secondly, the results of our analysis also shed light on the advantages and drawbacks of the conventional parameter estimation method (MCMC) versus the machine learning reconstruction technique (GP), as far as their applicability to next-generation GW missions' data is concerned. 

We find that while Fisher analysis always forecasts a higher tension due to tighter constraints on $H_0$ while keeping the mean fixed, MCMC predicts relatively shifted mean values resulting in relaxed tensions for models in higher tension with SH0ES, and no significant change for models closer to the local measurement of $H_0$. GP reveals shifts in the mean values of the reconstructed $H_0$ to higher values, which gives rise to relaxed tension when assuming fiducials which are in higher tension with SH0ES. There is hardly any noticeable change in the reconstructed mean values for fiducials lying closer to the SH0ES estimate. Finally, we suggest that any comment regarding the Hubble tension for a future mission should be carried out by employing all three methods in order to make the analysis robust and the conclusions more concrete, until the community is sure about a strong and competitive advantage of a particular approach over the others.

The plan of the paper is as follows. In Sec. \ref{sec:models}, we briefly discuss the six representative classes of cosmological models we choose for the purpose of this study. In Sec. \ref{sec:latestdata}, we find out the updated constraints on the parameters of each model based on the latest available datasets of CMB + BAO + SNIa. In particular, we highlight the constraints on four models for which we perform our own MCMC analysis due to inadequacies of previous analyses in the literature, in order to bring all the models to an equal footing.  Section \ref{sec:catalogue} deals with the outline of our adopted procedure to generate mock eLISA catalogs utilizing fiducial values based on the present constraints from the chosen models. In Sec. \ref{sec:threepronged}, a three-pronged forecasting methodology is employed to analyze the prospects of the individual models, or fiducials from each class of models in addressing the Hubble tension in light of eLISA, as predicted by the parametric and non-parametric approaches. This constitutes a thorough comparative study of the merits and drawbacks of each method when used as a forecasting tool. In Sec. \ref{sec:analysis} and \ref{sec:conclusion}, we summarize our key findings and comment on future directions which warrant further exploration in future studies.


\section{\label{sec:models}Models/parametrizations under consideration}
As is well-known, the direct measurement of $H_0$ is done via the measurement of luminosity distance. For a spatially flat, homogeneous and isotropic Universe, the luminosity distance ($d_L$) is defined as
\begin{equation} \label{dL}
    d_{L}=\frac{c(1+z)}{H_{0}} \int_{0}^{z} \frac{d x}{E\left(x\right)}\:\:,
\end{equation}
where $H_0$ is the Hubble parameter at present epoch, \textit{i.e.}, the Hubble constant, and $E(z) = H(z)/H_0$ is the reduced Hubble parameter. On assuming general relativity (GR), the integrand above can be approximated with
\begin{equation} \label{E2}
    E^2(z) = {\Omega_{m0} (1+z)^{3}+ \Omega_{r0} (1+z)^4 + \left(1-\Omega_{m0} -\Omega_{r0}\right) \exp \left[ 3 \int_{0}^{z} \frac{1+w(x)}{1+x} d x\right] }\:.
\end{equation}
Here, $\Omega_{m0}$ is the matter density parameter at the present epoch, $\Omega_{r0}$ is the radiation density parameter at the present epoch, and $w(z)$ is the equation of state (EoS) of the dark energy (DE) sector which is assumed in general to be of dynamic nature.

While this definition of $d_L$ typically applies to electromagnetic sources (such as supernovae), within GR it also holds for GW sources as there is no distinction between the EM luminosity distance and the GW luminosity distance $d_L^{(GW)}$ \cite{modgrav1,modgrav2,modgrav3}. Since we have restricted our present analysis to GR, we shall henceforth identify $d_L$ with the GW luminosity distance throughout the rest of this paper. 

Generically, for a standard siren event, $d_L$ can be directly inferred from the waveform. The redshift can subsequently be inferred from an electromagnetic counterpart, or from cross-correlation with large scale structure (LSS) catalogs \cite{LIGO_LSS_1,LIGO_LSS_2,LIGO_H0}. Hence, astrophysical GW events at intermediate redshifts can be an efficient probe for constraining the background cosmological parameters which appear in \eqref{dL}. 

In this work, we have focused on a few representative class of cosmological models via parametrizations which have shown promise in alleviating the Hubble tension to various extents in light of currently available datasets. To keep the extended parameter space minimal, we have considered examples of only zero, one, and two parameter extensions to the baseline $\Lambda$CDM model. We also justify considering the particular models for our analysis. Extensions with more than two extra parameters have not been considered in this study, as such parametrizations tend to fare poorly in terms of model selection criteria, \textit{e.g.}, when subjected to Akaike Information Criterion (AIC) and/or Bayesian Information Criterion (BIC) tests \cite{info1,info2,info3}.

\subsection{6-parameter scenarios}

\subsubsection{\texorpdfstring{$\Lambda$ Cold Dark Matter ($\Lambda$CDM)}{}}
We include $\Lambda$CDM as the benchmark model in our study, as we are interested in comparing the performances of the alternative models against the baseline model.
Latest constraints from \textit{Planck} 2018 (based on joint analysis of TT+TE+EE+low E+lensing data) currently yield $H_0=67.36\pm0.54$ km s\textsuperscript{-1} Mpc\textsuperscript{-1} for $\Lambda$CDM  \cite{Pl18VI}. As pointed out earlier, this is almost in $5\sigma$ tension with the locally measured value of $H_0=73.30\pm1.04$ km s\textsuperscript{-1} Mpc\textsuperscript{-1} reported by R21 \cite{Riess_2022}. 

\subsubsection{Phenomenologically Emergent Dark Energy (PEDE)}
Introduced in \cite{Li_2019}, the Phenomenologically Emergent Dark Energy model proposes a time-varying dark energy density parameter of the form
\begin{equation}
    \tilde{\Omega}_{\text{PEDE}}(z)=\Omega_{\text{PEDE},0}\left[1-\textrm{tanh}(\textrm{log}_{10}(1+z))\right]\:.
\end{equation}
In this model, there is no significant DE contribution at early times (high $z$), while a redshift-dependent DE component gradually emerges at late times. The effective EoS of the DE fluid is phantom-like and has the form
\begin{equation}
    w(z)=-1-\dfrac{1}{3\textrm{ln}10}[1+\textrm{tanh}(\textrm{log}_{10}(1+z))]\:\:,
\end{equation}
which reduces to the $\Lambda$CDM value of $w_\Lambda=-1$ at the present epoch. Constraints for the PEDE model based on \textit{Planck} 2018 alone yield $H_0=72.35\pm0.78$ km s\textsuperscript{-1} Mpc\textsuperscript{-1} whereas a joint analysis with \textit{Planck} 2018 + CMB lensing + BAO
+ Pantheon + DES + R19 yields $H_0=72.16\pm0.44$ km s\textsuperscript{-1} Mpc\textsuperscript{-1}, both appearing to alleviate the tension with R20 within $1\sigma$ \cite{YANG2021100762}. However, an unbiased analysis of the $H_0$ tension with local measurements should not include any direct measurement prior when it comes to parameter estimation, as it induces an inherent bias towards higher values of $H_0$.  Also, by now R21 data is available, so the previous conclusions based on this model needs to be revisited. In the following section, we overcome this issue by performing our own MCMC analysis for this model with relevant datasets.

\subsubsection{Vacuum Metamorphosis (VM)}
The Vacuum Metamorphosis model, proposed originally to explain the late-time accelerated expansion of the Universe, is motivated by non-perturbative quantum gravitational effects in curved spacetime \cite{vm1,vm2,vm3}. It invokes a minimally coupled, ultra-light scalar field of mass $m\sim10^{-33}$ eV. The Ricci scalar acts as the order parameter for a gravitational phase transition, which occurs as $R$ drops down to the critical value of $\chi m^2$ around $z\sim1$ (with $\chi\sim1$ being a dimensionless parameter). Thereafter, the setup mimics a late-time accelerated scenario, while a vacuum feedback mechanism prevents $R$ from decreasing any further. The gravitational phase transition occurs at the critical redshift
\begin{equation}
z_{c}=-1+\dfrac{3\Omega_{m0}}{4(1-M)}\:\:,
\end{equation}
where $M=m^2/12H_0^2$. In the original VM model, $M$ is not an independent parameter but is related to the matter density via the relation
\begin{equation} \label{momeq}
    \Omega_{m0}=\dfrac{4}{3}[3M(1-M)^3]^{1/4}\:.
\end{equation}
Neglecting spatial curvature, cosmic expansion is governed by the equations
\begin{equation} \label{vmorigbefore}
    H^2/H_0^2=\Omega_{m0}(1+z)^3+\Omega_{r0}(1+z)^4\quad\textrm{for}\:\: z>z_{c}\:,
\end{equation}
\begin{equation} \label{vmorigafter}
    H^2/H_0^2=(1-M)(1+z)^4+M\quad\textrm{for}\:\: z<z_{c}\:.
\end{equation}
Joint analysis with \textit{Planck} 2018 + BAO + Pantheon yields $H_0=74.21\pm0.66$ km s\textsuperscript{-1} Mpc\textsuperscript{-1} \cite{VM_Di_Valentino_2020}, which somewhat overshoots the mean $H_0$ obtained from R21. It, nonetheless, resolves the tension with R21 within $1\sigma$. But as pointed out in \cite{VM_Di_Valentino_2020}, the success of the model in resolving the $H_0$ tension should not be viewed in isolation, as it suffers from a poorer goodness of fit to combined CMB + BAO + SNIa datasets compared to $\Lambda$CDM. So, this model is an interesting one that needs to be taken into account during our investigation in order to check its credentials against future observations.

\subsection{1-parameter extension}

\subsubsection{Elaborated Vacuum Metamorphosis (VM-VEV)}
This is an extended version of the original VM model, where the scalar field is allowed to have a non-zero vacuum expectation value (VEV) which shows up as a cosmological constant at $z>z_{c}$. While cosmic evolution after the phase transition remains identical to \eqref{vmorigafter}, the pre-transition history is modified to
\begin{equation} \label{evmbefore}
    H^2/H_0^2 = \Omega_{m0}(1+z)^3+\Omega_{r0}(1+z)^4  +M\Bigg[1-\left\lbrace 3\left(\dfrac{4}{3\Omega_{m0}}\right)^4M(1-M)^3\right\rbrace^{-1}\Bigg]\:.
\end{equation}
Here, $M$ is no longer related to $\Omega_{m0}$ as in \eqref{momeq} but appears as a free parameter. The VM-VEV model, therefore, allows a non-vanishing dark energy component even before the gravitational phase transition has taken place. Furthermore, in order to ensure $z_{c}\geq0$ and $\Omega_{DE}(z>z_{c})\geq0$, theoretical limits have to be placed on the prior as
\begin{equation}
    \dfrac{4}{3}(1-M)\leq\Omega_{m0}\leq\dfrac{4}{3}\left[3M(1-M)^3\right]^{1/4}\:.
\end{equation}
Joint analysis with \textit{Planck} 2018 + BAO + Pantheon leads to $H_0=73.26\pm0.32$ km s\textsuperscript{-1} Mpc\textsuperscript{-1} \cite{VM_Di_Valentino_2020}, which is in better agreement with R21 than the original VM model. The goodness of fit is somewhat improved compared to VM, although it still remains worse compared to $\Lambda$CDM. So, we have considered this model from a similar point of view as for VM.

\subsection{2-parameter extensions}

\subsubsection{Chevallier-Polarski-Linder (CPL)}
The CPL parametrization \cite{cpl1,cpl2}, also known as the $w_0w_a$CDM parametrization, is perhaps the most widely used model after $\Lambda$CDM. It is a two-parameter extension to $\Lambda$CDM with a redshift-dependent DE EoS given by
\begin{equation}
    w(z)=w_0+w_a\frac{z}{1+z}\:\:.
\end{equation}
It can be interpreted as the first-order Taylor expansion of a more generic DE EoS $w(z)$ in terms of the scale factor $a=(1+z)^{-1}$ \cite{cpltaylor}, that is well-behaved at both high and low redshifts. The present value of the EoS is given by $w_0$, while it also remains bounded by $(w_0+w_a)$ in the far past. Its simple form and its ability to parameterize a wide range of theoretical DE models render it a particularly appealing choice. Joint analysis with \textit{Planck} 2018 + BAO + Pantheon in \cite{Pl18VI} yields $H_0=68.31\pm0.82$ km s\textsuperscript{-1} Mpc\textsuperscript{-1}, which alleviates the tension with R21 to approximately $3.8\sigma$. It is quite natural to investigate the prospects of this widely used model against future GW data.

\subsubsection{Jassal-Bagla-Padmanabhan (JBP)}
More generally, one can proceed to construct a class of CPL-like parametrizations where the redshift-dependence scales as $z/(1+z)^p$, with $p$ being a natural number. The JBP parametrization \cite{jbp} is the next example in this family with $p=2$, \textit{i.e.}, it proposes a DE EoS of the form
\begin{equation}
    w(z)=w_0+w_a\frac{z}{(1+z)^2}\:\:.
\end{equation}
When constrained with \textit{Planck} 2018 + BAO alone, the JBP parametrization has been shown to admit $H_0=67.4^{+1.9}_{-2.9}$ km s\textsuperscript{-1} Mpc\textsuperscript{-1}, which relaxes the tension with R20 down to $2.7\sigma$ \cite{Yang2021}. However, one should take this result with a pinch of salt as it is partly due to larger error bars that result from the absence of SNIa data in the analysis. We remind the reader that while addressing $H_0$ tension from a particular model and comparing it with other models with similar targets, one should take into account SNIa data as well. In this work, we have brought JBP to an equal footing with its peers by including SNIa data while re-estimating its parameters. 


\section{\label{sec:latestdata}Updated constraints from latest datasets}

As noted in the previous section, even though there are series of models that aim to address the $H_0$ tension from different perspectives, a consistent and methodical analysis was found somewhat lacking in the literature, at least for a number of interesting models that we have taken into consideration. To summarize, this was found to be the case in either of the following two ways:
\begin{enumerate}[(1)]
    \item When one aims to study the tension between two distinct datasets (\textit{e.g.}, the value of $H_0$ obtained from \textit{Planck} 2018 against that from R21), one should not combine those datasets, jointly estimate the value of a model parameter, and then compare it with the standalone result from one of the datasets used in the process of estimation (\textit{e.g.}, with R21). This circular procedure defeats the purpose itself and leads to a biased analysis, as done in a good fraction of previous works. More often than not, it mistakenly leads to a smaller estimate of the tension than what actually exists between the datasets.
    \item A few of the previous attempts to address the $H_0$ tension with the cosmological models chosen here either do not consider the latest datasets, or miss one/more of the important datasets (\textit{e.g.}, SNIa). While the former simply results in outdated constraints on model parameters, the latter is a more serious issue and may lead to a scientifically inaccurate analysis with questionable conclusions. As far as the latest status of the $H_0$ tension is concerned, neither category of works can be completely relied upon.
\end{enumerate}

\begin{table*}[!t]
    \begin{minipage}{0.5\linewidth}
        \begin{center}
            {\renewcommand{\arraystretch}{1.1} \setlength{\tabcolsep}{20 pt} \centering  
            \begin{tabular}{|c|c|}
                \hline
                \textbf{Parameter}             & \textbf{Prior}\\
                \hline
                $\Omega_{\rm b} h^2$           & $[0.005,0.1]$\\
                $\Omega_{\rm c} h^2$           & $[0.01, 0.99]$\\
                $100\theta_{s}$                & $[0.5,10]$\\
                $\ln\left(10^{10}A_{s}\right)$ & $[1,4]$\\
                $n_{s}$                        & $[0.5, 1.5]$\\
                $\tau$                         & $[0.005,0.8]$\\
                $w_0$                          & $[-2, 1]$\\
                $w_a$                          & $[-3, 3]$\\
                $M$*                           & $[0.5, 1]$\\
                \hline
            \end{tabular}
            }
            \caption{Uniform priors on cosmological model parameters (*quoted from \cite{VM_Di_Valentino_2020})}
            \label{tab:priors}
        \end{center}
    \end{minipage}\hfill
    \begin{minipage}{0.48\linewidth}
        \begin{center}
            \includegraphics[width=0.95\textwidth]{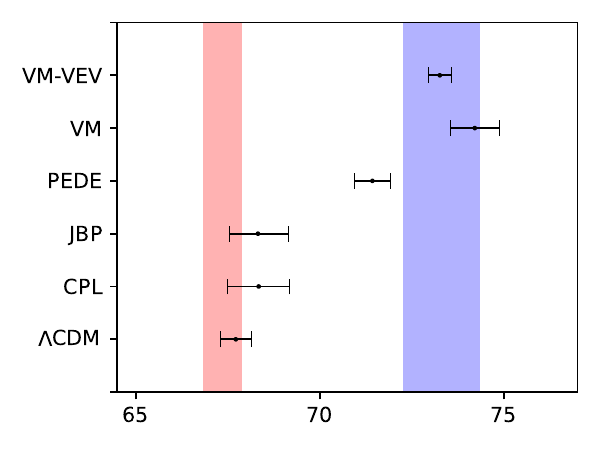}
            \captionof{figure}{Latest constraints on $H_0$ (in units of km s\textsuperscript{-1} Mpc\textsuperscript{-1}) from the \textit{Planck} 2018 + BAO + Pantheon MCMC analyses.}
            \label{currentwhisker}
        \end{center}
    \end{minipage}
\end{table*}

\begin{figure*}[!h]
\begin{center}
    \begin{subfigure}{.48\textwidth}
        \includegraphics[width=\textwidth]{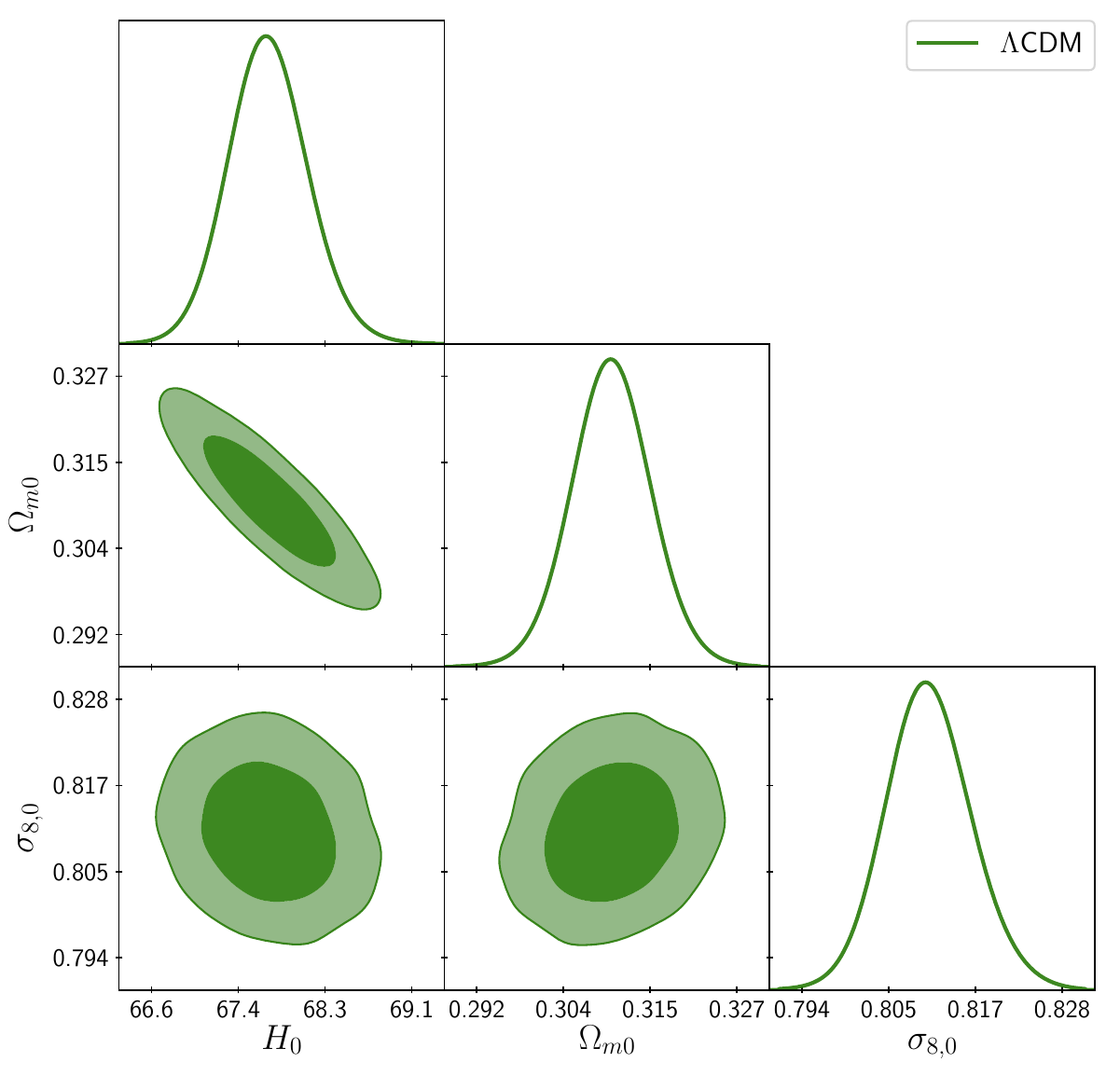}
        \caption{$\Lambda$CDM}
    \end{subfigure}
    \begin{subfigure}{.48\textwidth}
        \includegraphics[width=\textwidth]{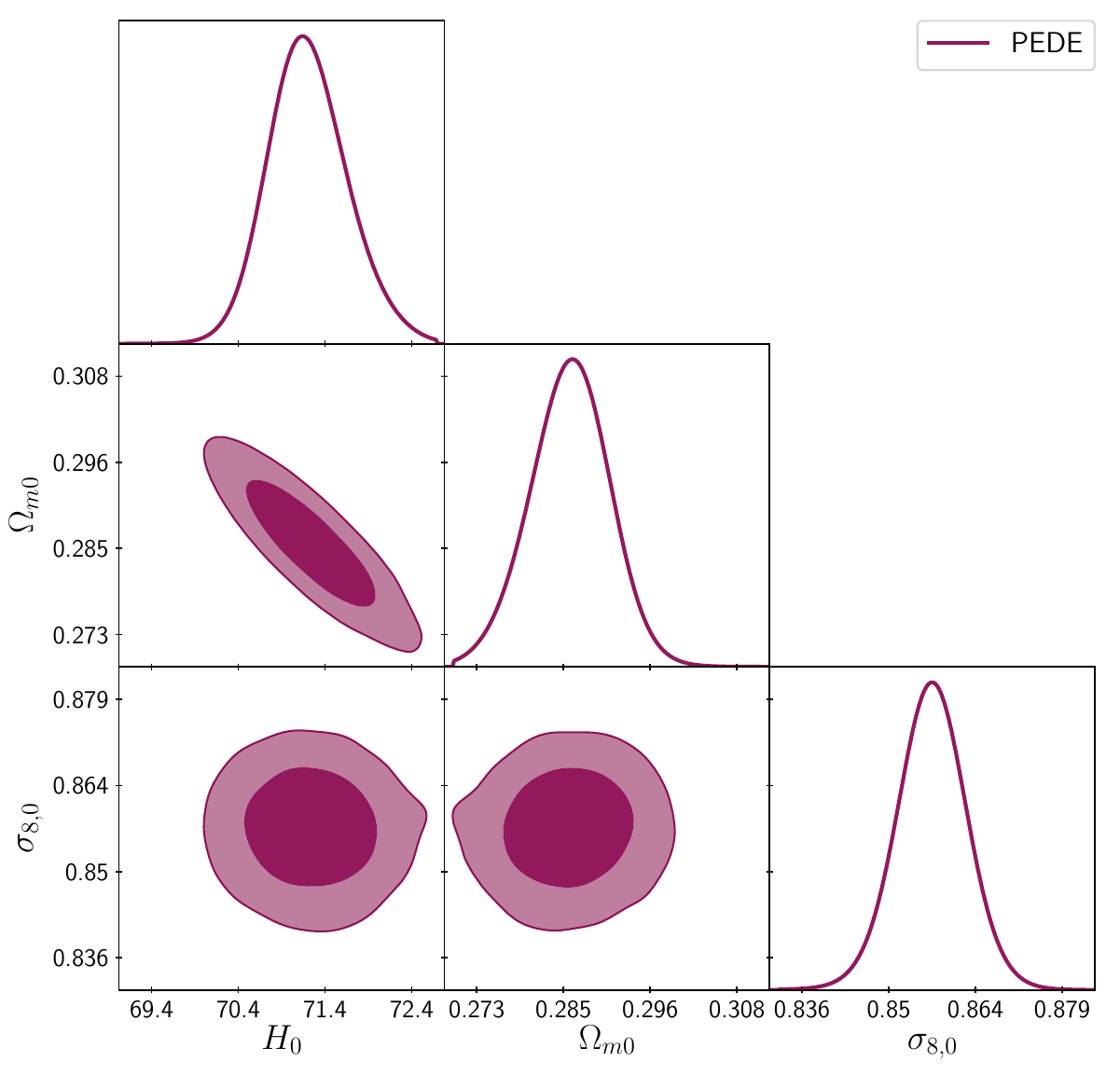}
        \caption{PEDE}
    \end{subfigure}
    \bigskip
    \begin{subfigure}{.485\textwidth}
        \includegraphics[width=\textwidth]{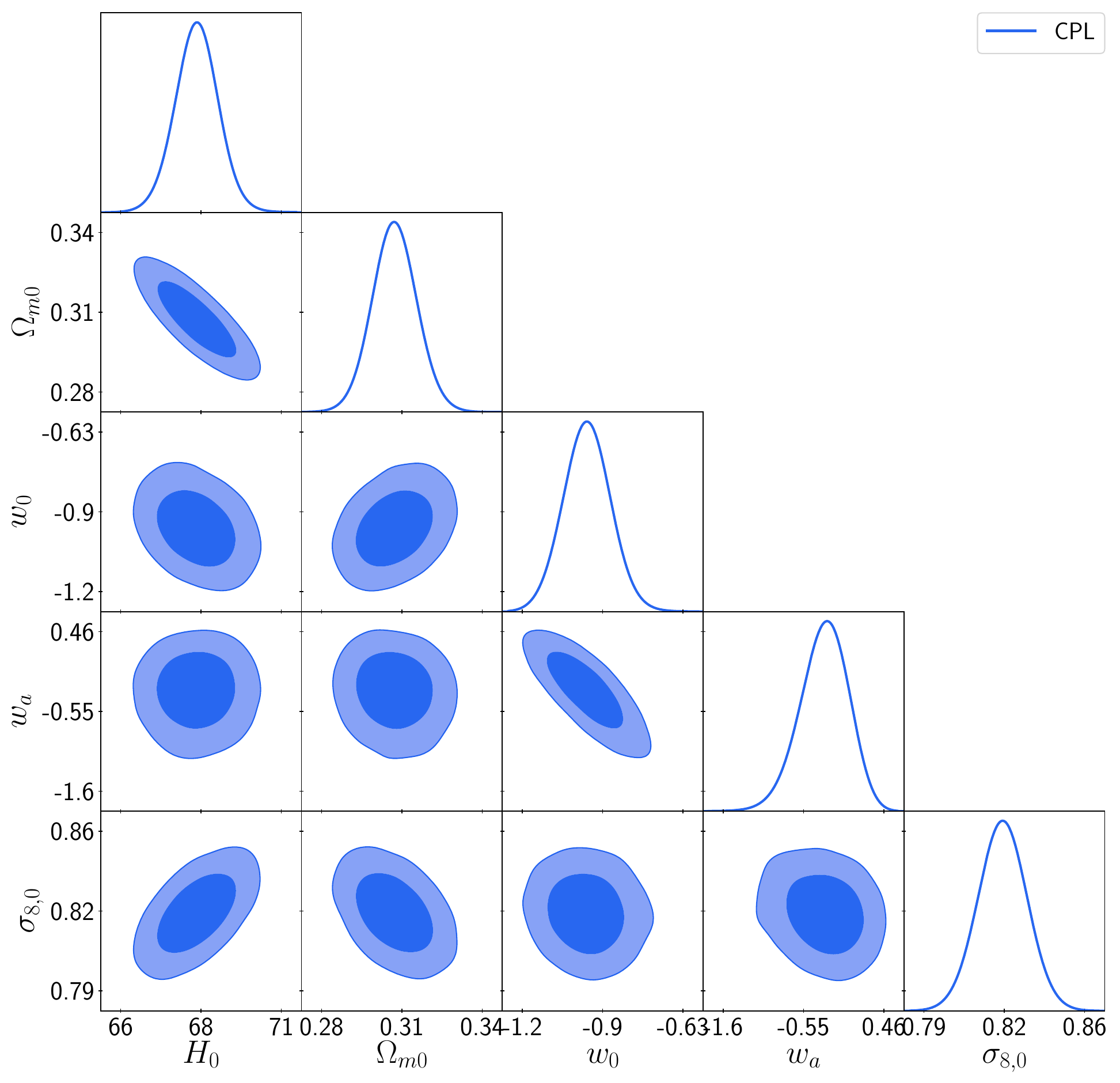}
        \caption{CPL}
    \end{subfigure}
    \begin{subfigure}{.485\textwidth}
        \includegraphics[width=\textwidth]{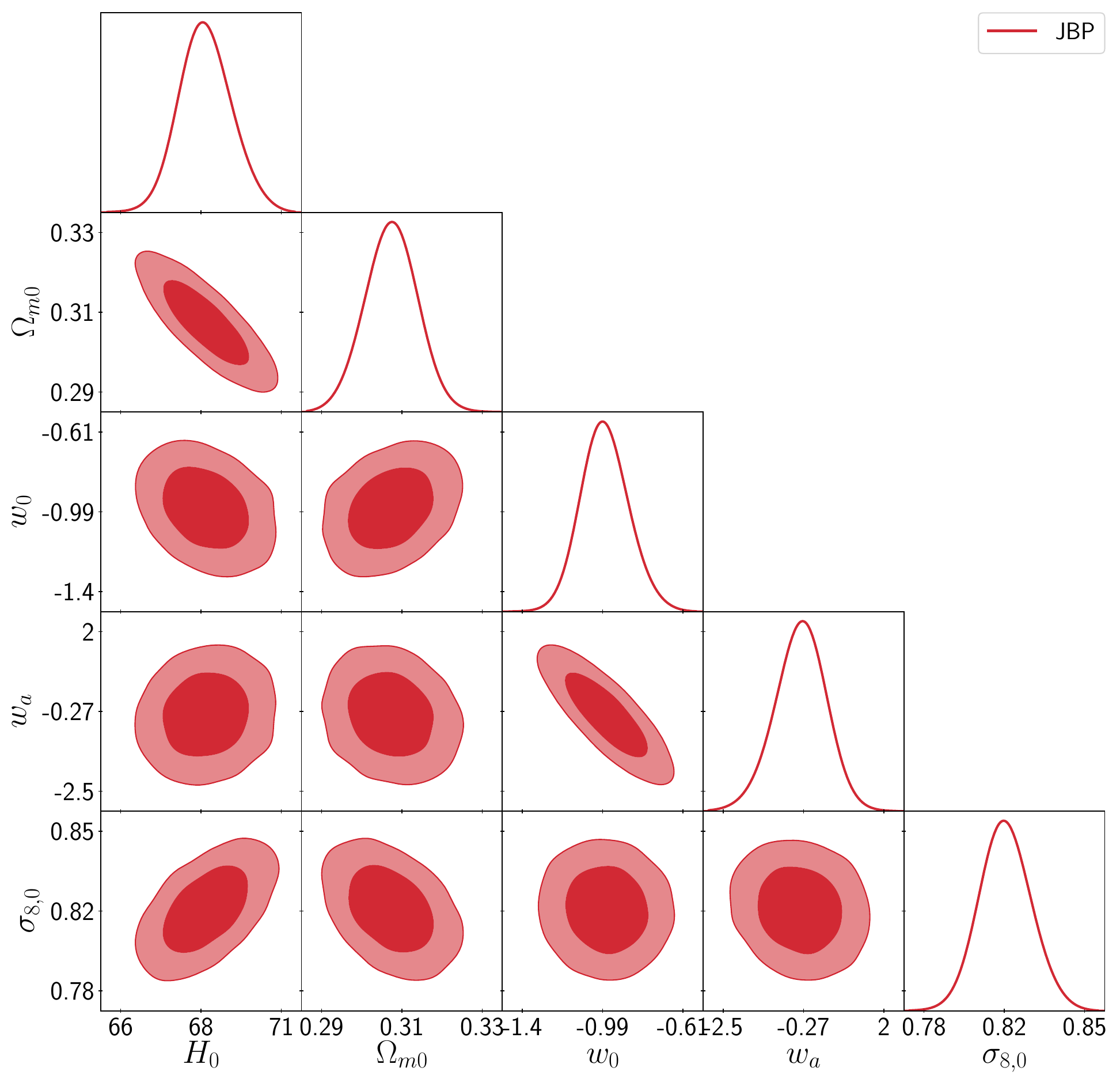}
        \caption{JBP}
    \end{subfigure}
\end{center}
    \vspace*{-0.8cm}
    \caption{Markov Chain Monte Carlo contours using latest datasets \textit{Planck} 2018 + BAO + Pantheon. For VM and VM-VEV please refer to \cite{VM_Di_Valentino_2020}.}
    \label{fig:mcmccurrent}
\end{figure*}

In order to put all the models on a uniform footing and estimate their latest tensions with direct measurements of $H_0$ from R21, we reanalyze the models associated with unsatisfactory analyses  with the same set of latest cosmological datasets (by combining the latest CMB + BAO + SNIa datasets) and constraining their parameter spaces by MCMC. We have also kept direct measurement data outside this MCMC for the reason mentioned above. These constraints are later utilized in Sec \ref{sec:catalogue} as fiducials to generate the synthetic catalogs for forecasting. So, this exercise will enable us to investigate the prospects of future missions in a thorough and unbiased fashion and inspect the role played by these well-motivated fiducial values. It would also help us make an honest comparison among the different models under consideration. 

We consider the following datasets for constraining our chosen models in a consistent manner:

\begin{enumerate}[(1)]
    \item \textbf{CMB}: Cosmic Microwave Background temperature and polarization angular power spectra, and CMB lensing of \textit{Planck} 2018 \textit{TTTEEE+low l+low E+lensing} \cite{Pl18V,Pl18VI,Pl18VIII}.
    \item \textbf{BAO}: Baryon Acoustic Oscillations measurements by 6dFGS \cite{6dF}, SDSS MGS \cite{MGS}, and BOSS DR12 \cite{BOSS} (as used in the \textit{Planck} 2018 analysis \cite{Pl18VI}).
    \item \textbf{SNIa}: Luminosity distance data of 1048 type Ia supernovae from the Pantheon compilation \cite{Pantheon}.
\end{enumerate}
Additionally, we have used the Riess \textit{et al.} (2021) dataset \cite{Riess_2022} for comparison of the value of $H_0$ with that obtained from the chosen models. However, we have not included this dataset during parameter estimation, because of the reason mentioned earlier.

\begin{table*}[!ht]
    \resizebox{1.0\textwidth}{!}{\renewcommand{\arraystretch}{1.65} \setlength{\tabcolsep}{10 pt} 
    \begin{tabular}{c c c c c c c c}
        \hline\hline
        \textbf{Parameters} & $\boldsymbol{\Lambda}$\textbf{CDM} & \textbf{CPL} & \textbf{JBP} & \textbf{PEDE} & \textbf{VM}* & \textbf{VM-VEV}* \\ \hline
        $\Omega_b h^2$ & $0.02243_{-0.00014}^{+0.00013}$ & $0.02238_{-0.00015}^{+0.00014}$ & $0.02239_{-0.00014}^{+0.00015}$ & $0.02222_{-0.00013}^{+0.00013}$ & $0.02213_{-0.00012}^{+0.00012}$ & $0.02236_{-0.00015}^{+0.00015}$ \\
        $\Omega_c h^2$ & $0.1192_{-0.00095}^{+0.00091}$ & $0.12_{-0.0011}^{+0.0011}$ & $0.1198_{-0.0011}^{+0.0011}$ & $0.122_{-0.00089}^{+0.00088}$ & $-$ & $0.1217_{-0.0012}^{+0.0012}$  \\
        $100\theta_{s}$ & $1.042_{-0.00028}^{+0.00029}$ & $1.042_{-0.00029}^{+0.00031}$ & $1.042_{-0.00028}^{+0.0003}$ & $1.042_{-0.0003}^{+0.00029}$ & $1.04053_{-0.00029}^{+0.00029}$ & $1.04077_{-0.00030}^{+0.00030}$  \\
        $\ln\left(10^{10}A_{s}\right)$ & $3.048_{-0.014}^{+0.014}$ & $3.043_{-0.015}^{+0.015}$ & $3.043_{-0.015}^{+0.015}$ & $3.033_{-0.013}^{+0.014}$ & $3.035_{-0.014}^{+0.017}$ & $3.044_{-0.014}^{+0.016}$  \\
        $n_{s}$ & $0.9672_{-0.0038}^{+0.0037}$ & $0.9653_{-0.0041}^{+0.0039}$ & $0.9658_{-0.0042}^{+0.0038}$ & $0.9604_{-0.0037}^{+0.0037}$ & $0.9648_{-0.0043}^{+0.0043}$ & $0.9636_{-0.0045}^{+0.0045}$  \\
        $\tau$ & $0.05682_{-0.0074}^{+0.0069}$ & $0.05348_{-0.0079}^{+0.0075}$ & $0.05362_{-0.0078}^{+0.0075}$ & $0.04685_{-0.0069}^{+0.0075}$ & $0.0483_{-0.0067}^{+0.0079}$ & $0.0528_{-0.0077}^{+0.0077}$   \\
        $w_0$ & $-$ & $-0.9571_{-0.082}^{+0.078}$ & $-0.9705_{-0.12}^{+0.12}$ & $-$ & $-$ & $-$  \\
        $w_a$ & $-$ & $-0.2904_{-0.28}^{+0.33}$ & $-0.3648_{-0.78}^{+0.74}$ & $-$ & $-$ & $-$  \\
        $M$ & $-$ & $-$ & $-$ & $-$ & $0.9277_{-0.0044}^{+0.0044}$ & $0.8929_{-0.0016}^{+0.0010}$  \\
        \hline 
        $H_0$ & $67.72_{-0.41}^{+0.42}$ & $68.34_{-0.85}^{+0.83}$ & $68.32_{-0.82}^{+0.78}$ & $71.24_{-0.48}^{+0.49}$ & $74.21_{-0.66}^{+0.66}$ & $73.26_{-0.32}^{+0.32}$  \\
        $\Omega_{m0}$ & $0.3102_{-0.0057}^{+0.0054}$ & $0.3064_{-0.0081}^{+0.0079}$ & $0.3062_{-0.0078}^{+0.0075}$ & $0.2855_{-0.0056}^{+0.0051}$ & $0.2593_{-0.0046}^{+0.0046}$ & $0.2695_{-0.0041}^{+0.0041}$  \\
        $\sigma_{8,0}$ & $0.8105_{-0.0059}^{+0.0059}$ & $0.8208_{-0.011}^{+0.011}$ & $0.8185_{-0.011}^{+0.011}$ & $0.8572_{-0.0061}^{+0.0064}$ & $0.9461_{-0.0068}^{+0.0080}$ & $0.8756_{-0.0091}^{+0.0091}$  \\
        \hline
        \hline
    \end{tabular}
    }
\caption{Latest constraints on the parameters of the cosmological models considered in Sec. \ref{sec:models} using combined \textit{Planck} 2018 + BAO + Pantheon observational data (constraints marked with * have been quoted from \cite{VM_Di_Valentino_2020}).}
\label{constraints}
\end{table*}

The priors used for the MCMC analyses using \textit{CLASS} \cite{CLASS1,CLASS2} and \textit{MontePython} \cite{MontePython1,MontePython2} are given in Table \ref{tab:priors}. Relevant modifications have been made in the codes at the background level only, with the standard CDM perturbation sector left unaltered as we do not consider the presence of any extra perturbed species. We have generated and analyzed the MCMC chains for $\Lambda$CDM, CPL, JBP, and PEDE. Constraints for the VM and VM-VEV models have been quoted from \cite{VM_Di_Valentino_2020} as their datasets used are consistent with ours. All the constraints are presented in Table \ref{constraints}. The status of the different models and parametrizations with respect to the Hubble tension in light of the latest datasets are summarized in the whisker plot shown in Fig. \ref{currentwhisker}. The blue bar indicates the latest SH0ES constraint \cite{Riess_2022}, and the red bar, the \textit{Planck} 2018 constraint \cite{Pl18VI} on the Hubble constant.

Figure \ref{fig:mcmccurrent} shows the outcome of MCMC analysis from latest datasets for the models under consideration. MCMC results for two of the chosen models, in particular, should be highlighted. Firstly, our joint CMB + BAO + SNIa analysis shows that the PEDE model is capable of reducing tension with the latest R21 measurement of $H_0$ down to $1.8\sigma$. However, it cannot resolve the tension within $1\sigma$, contrary to the claim of \cite{YANG2021100762}. Secondly, for the JBP model, our analysis yields a value of $H_0=68.32^{+0.78}_{-0.82}$ km s\textsuperscript{-1} Mpc\textsuperscript{-1} which is in roughly $3.8\sigma$ tension with R21. In other words, when confronted with SNIa data alongside CMB and BAO, the JBP parametrization does not offer any significant improvement over CPL ($\sim3.7\sigma$) when it comes to addressing the $H_0$ tension, unlike what has been concluded in \cite{Yang2021}. In fact, JBP is found to be marginally worse in performance compared to CPL. 

In what follows, we shall make use of these constraints as fiducials while generating mock eLISA catalogs for our analysis. As argued, this will lead to an up-to-date and scientifically accurate estimation of the relevant parameters from future missions in regard to the current datasets. 


\section{\label{sec:catalogue}Mock catalog generation}

Based on competing models of massive black hole formation, the MBHB population at intermediate redshifts detectable by eLISA can be broadly classified into three distinct source types as follows \cite{eLISA1,eLISA3}:
\begin{enumerate}[(1)]
\item \textbf{Pop III:} This is a light seed scenario where it is assumed that the first massive black holes grow from stellar remnants of early population III stars, which formed around $z\sim15-20$ within massive dark matter halos \cite{pop31,pop32}.
\item \textbf{No Delay:} This is a heavy seed scenario in which protogalactic disk collapse (\textit{e.g.}, due to bar instabilities) leads to the formation of massive black holes \cite{heavyseed1,heavyseed2}. The MBHB coalescence events are assumed to take place simultaneously with the mergers of their host galaxies, which is a simplistic premise of this model. 
\item \textbf{Delay:} This is a more realistic heavy seed scenario, in which there is a finite time delay between the merger of a given pair of host galaxies and that of the black holes. The intermediate period leading up to the MBHB merger is governed by a variety of complicated astrophysical processes \cite{delay1,delay2,delay3,delay4} which non-trivially affect the observable redshift distribution of the MBHB population. 
\end{enumerate}
Throughout this study, we have chosen to work with the L6A2M5N2 configuration of eLISA, which closely resembles the proposed mission specification \cite{eLISA,eLISA1,eLISA2,eLISA3}. As noted earlier, we choose to work with MHBH merger events which are expected to be accompanied by observable electromagnetic counterparts. For each source type, we generate a set of standard siren mock catalogs by considering each of the chosen cosmological models (described in Sec. \ref{sec:models}) as an individual background model in order to get unbiased analyses. This is in contrast to most approaches in the literature, where the mock catalog is almost always generated from  $\Lambda$CDM even though the model under consideration is different, that naturally leads to a bias in the analysis. In this work we get rid of this bias. Our catalog generation process is inspired by the method outlined in Sec. 3.2 of \cite{Ferreira}, which is based on the redshift distributions of the three source types summarized in \cite{Early_Int_DE}. Following that prescription, we proceed as follows\footnote{The codes may be made available upon reasonable request.}: 
\begin{enumerate}[(1)]
    \item Sample from the theoretical redshift distribution of MBHB events based on the particular mission's specifications to get the set of event redshifts.
    \item Compute the theoretical luminosity distance $d_L(z)$ at these redshifts assuming a particular cosmological model. While doing so, instead of merely using the mean values of the parameters, randomly sample from the Gaussian priors on the parameters obtained earlier from constraints using latest datasets (see Sec. \ref{sec:latestdata}). This has been done because the mean values from an MCMC analysis are not the only numbers that are physically relevant. Rather, every number within a $1\sigma$ bound of each mean value ought to be realistically considered.
    \item Consider the various sources of error in the measurement of $d_L(z)$ in light of the mission's specifications, and compute the total error $\Delta d_L(z)$.
    \item Sample the final $d_L(z)$ from a normal distribution by considering the theoretical $d_L(z)$ as the mean and the error computed in the previous step as the $1\sigma$ bound.
\end{enumerate}
This provides us with a set of catalogs, each of which contains a set of event redshifts $\{z\}$, the corresponding luminosity distances $\{d_L(z)\}$, and the observational errors in determining the latter $\{\Delta d_L(z)\}$. The Gaussian sampling in both step 2 and step 4 ensure that we obtain catalogs as realistically as possible, such that the values of the physical parameters for any given catalog do not coincide with the corresponding fiducial values used in the generation process.

We have generated 500 mock catalogs for each of the three distinct source types of MBHBs visible to eLISA, for each of the mission durations (5, 10, and 15 years). This exercise has been repeated separately for fiducials corresponding to each of the six cosmological models discussed in Sec. \ref{sec:models}. We then employ both parametric and non-parametric approaches on these catalogs and analyze the results thus obtained, as described in the following sections. 


\section{\label{sec:threepronged} Different approaches and results}

With the mock data in hand, there are different ways to proceed further. The usual and widely used methods of addressing the problem fall under the category of parametric approaches. Apart from the usual methods, there can also be alternative, non-parametric approaches to the issue of the Hubble tension using simulated GW data (this is also applicable, in principle, to any such future data). Such techniques, in general, go by the name of reconstruction tools. In what follows, we shall make use of two parametric methods, namely Fisher forecast and MCMC, and one non-parametric method called Gaussian Processes.

\subsection{Parametric methods}

\subsubsection{\label{sec:fisher}Fisher Forecast}

In the first approach we employ a straightforward Fisher matrix analysis in order to forecast on the behaviors of the cosmological parameters of the various models when subjected to eLISA data. For observations of the GW luminosity distance ($d_L$) and redshift ($z$), the Fisher matrix is given as \cite{Dodelson:2003ft}
\begin{equation}
    F_{i j}=\sum_{n=\{z\}} \frac{1}{\sigma_{n}^{2}} \frac{\partial d_{L}\left(z_{n}\right)}{\partial \theta_{i}} \frac{\partial d_{L}\left(z_{n}\right)}{\partial \theta_{j}}\:\:.
\end{equation}
We have assumed the parameters to be independent of their covariances, \textit{i.e.}, uncorrelated errors. Here, $F_{ij}$ is the ${ij}^{th}$ element of the Fisher matrix, $\{\theta_i\}$ is the set of parameters whose errors are to be determined in the context of eLISA, and $\sigma_n$ is the error in observation of $d_L$ at redshift $z_n$. The summation runs over the redshift distribution $n=\{z\}$, which contains the redshift points at which the Fisher matrix needs to be evaluated. The prior on the $i^{th}$ parameter is added as $(\sigma^{(i)}_{prior})^{-2}$ to the corresponding diagonal element $F_{ii}$, which gives the augmented matrix $\widetilde{F}_{ij}$. Finally, the inverse of $\widetilde{F}_{ij}$ gives the covariance matrix, and the square root of each diagonal element of the covariance matrix, \textit{i.e.}, $\sqrt{(\widetilde{F}^{-1})_{ii}}$, gives the required $1\sigma$ error forecast on the corresponding parameter $\theta_i$ \cite{Tegmark_1997}.

\begin{figure}[!h]
    \centering
    \includegraphics[width=\textwidth,height=0.43\textheight,keepaspectratio]{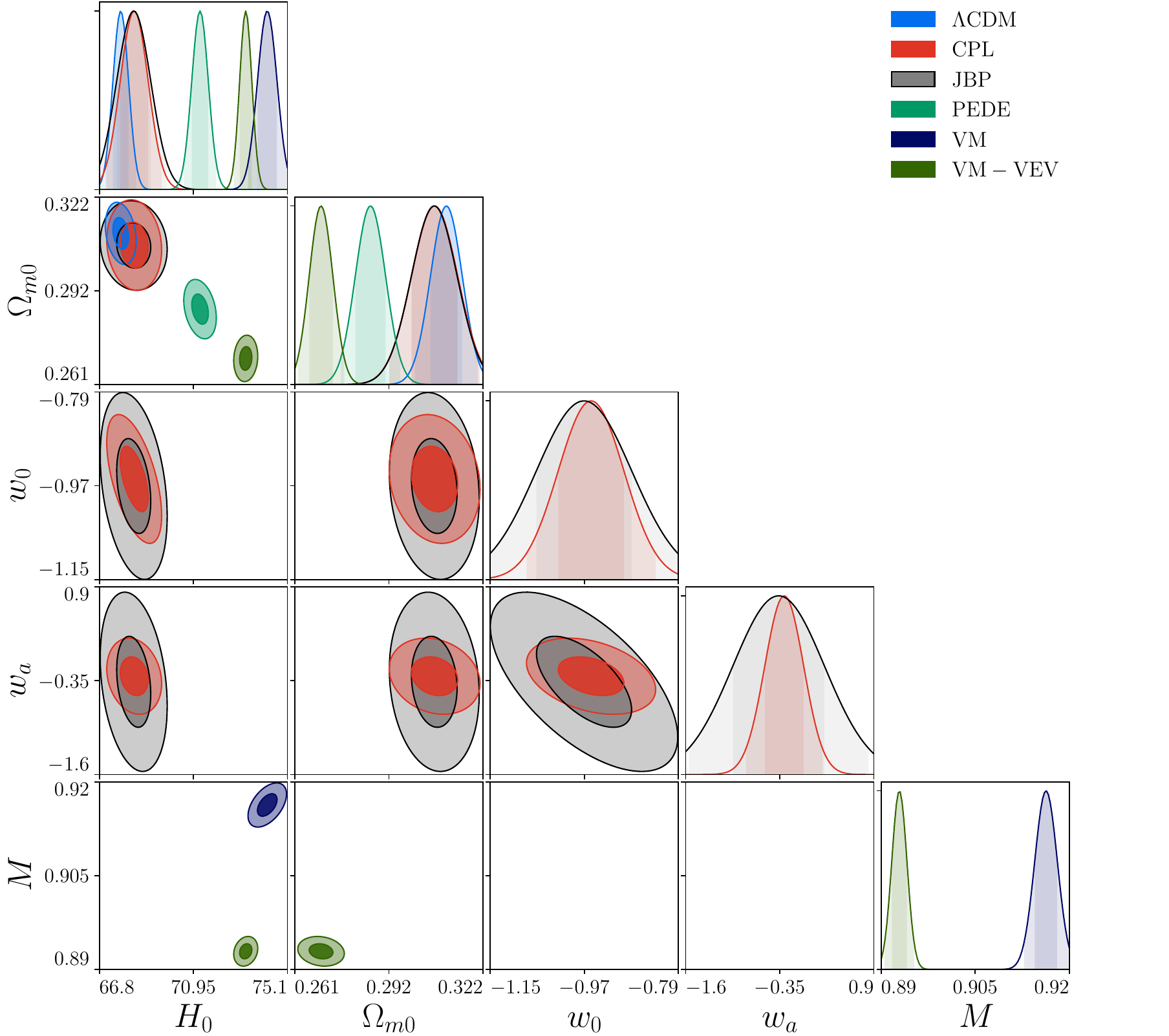}
    \caption{Forecast on 10 year eLISA mission duration for source type Delay.}
    \label{fig:fisherdelay}
\end{figure}

\begin{figure}[!h]
    \centering
    \includegraphics[width=\textwidth,height=0.43\textheight,keepaspectratio]{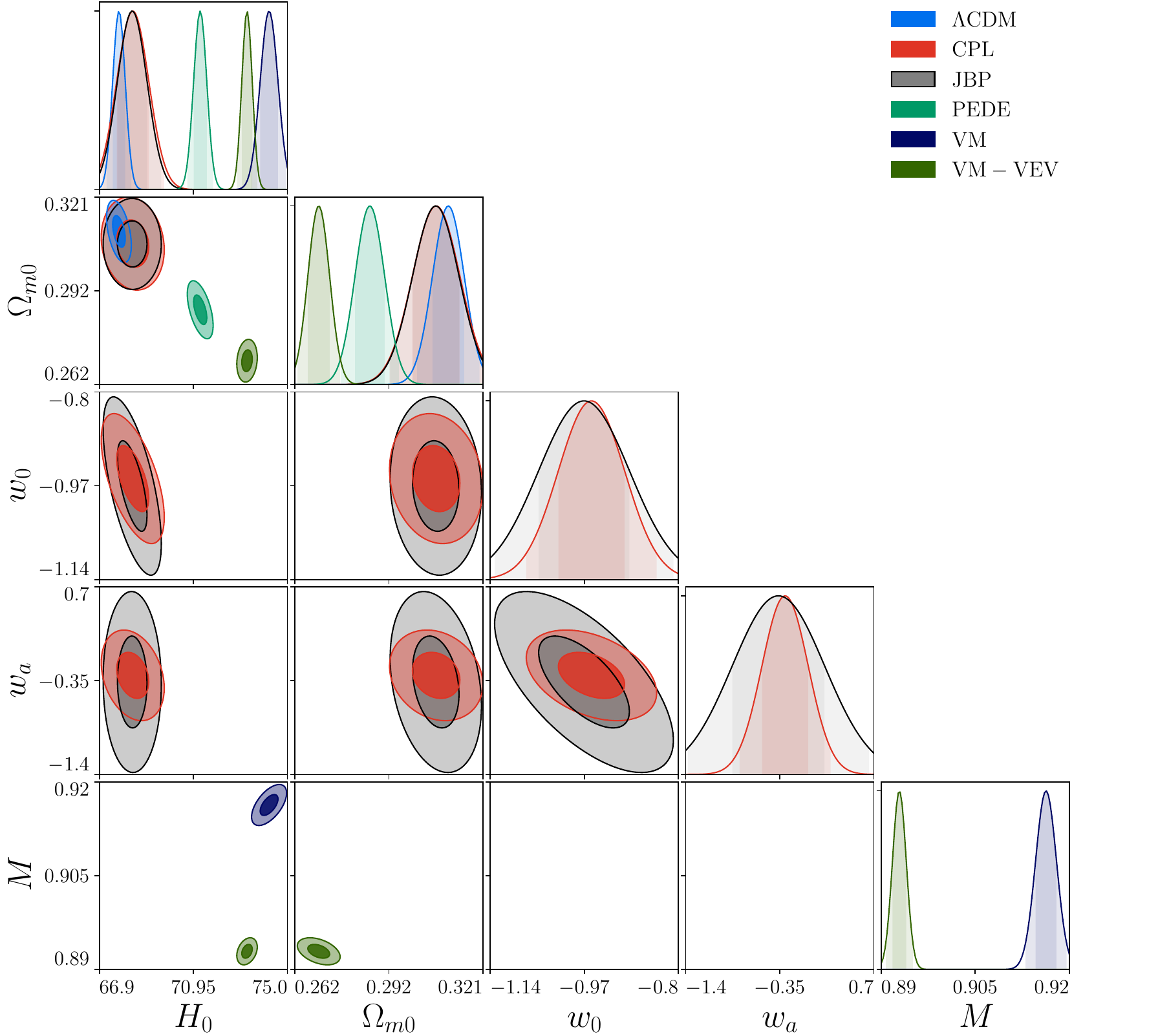}
    \caption{Forecast on 10 year eLISA mission duration for source type No Delay.}
    \label{fig:fishernodelay}
\end{figure}

\begin{figure}[!h]
    \centering
    \includegraphics[width=\textwidth,height=0.43\textheight,keepaspectratio]{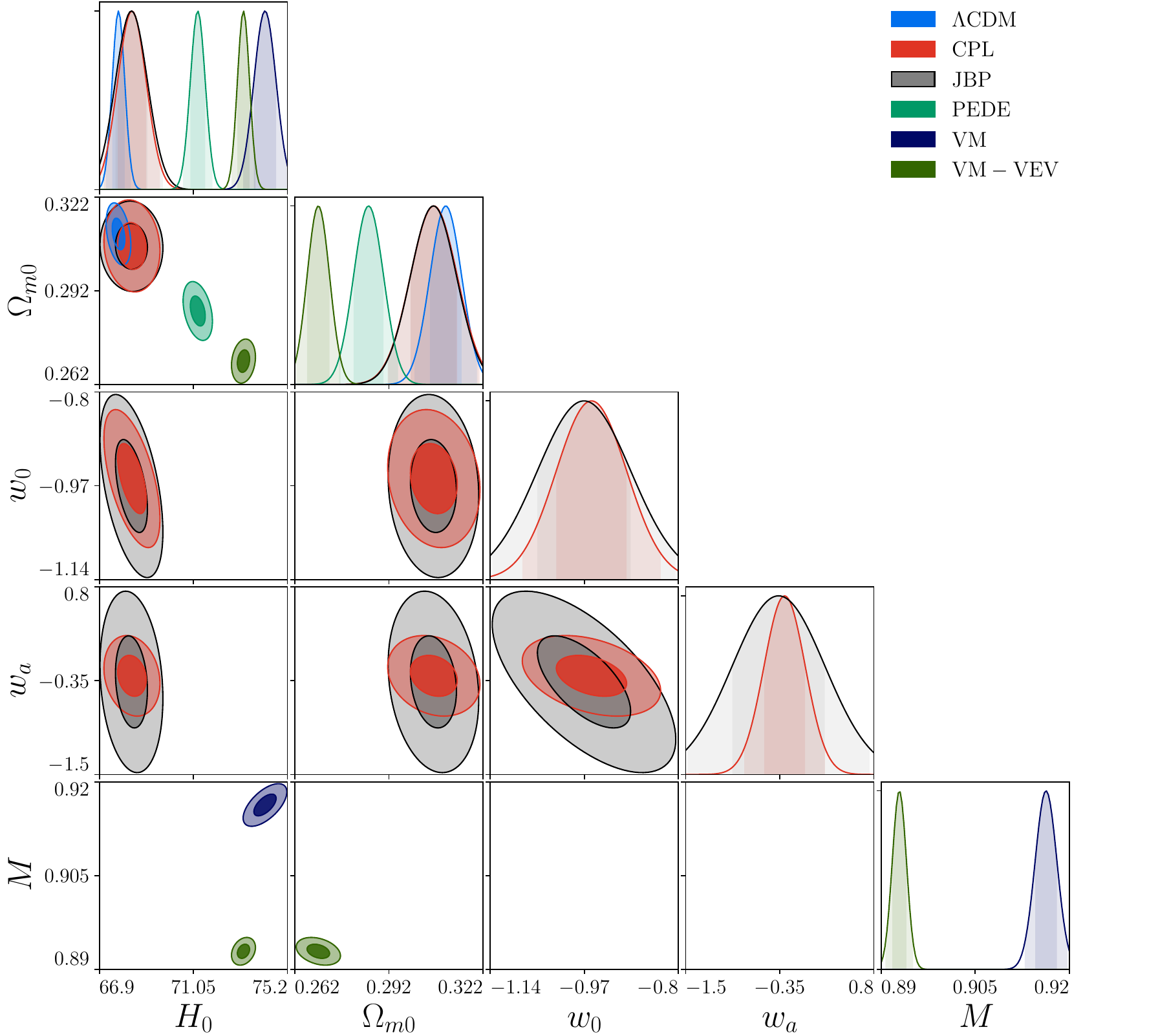}
    \caption{Forecast on 10 year eLISA mission duration for source type Pop III.}
    \label{fig:fisherpopIII}
\end{figure}

\begin{table}[!h]
    \centering
    \resizebox{\textwidth}{!}{\renewcommand{\arraystretch}{1.0} \setlength{\tabcolsep}{30 pt} 
    \begin{tabular}{c c c : c c c c c}
        \hline
        \textbf{Model} & \textbf{Source Type} & \textbf{Years} & \multicolumn{2}{c}{$\bm{\Delta H_0}$} & \multicolumn{2}{c}{$\bm{\Delta \Omega_m}$}\\
        \hline
        \multirow{9}{0.5cm}{LCDM} & \multirow{3}{0.5cm}{Delay} & 5 & \multicolumn{2}{c}{0.3509} & \multicolumn{2}{c}{0.0053} \\ 
        ~ & ~ & 10 & \multicolumn{2}{c}{0.3314} & \multicolumn{2}{c}{0.0051} \\ 
        ~ & ~ & 15 & \multicolumn{2}{c}{0.3033} & \multicolumn{2}{c}{0.0050} \\ \cdashline{2-7}
        ~ & \multirow{3}{0.5cm}{No Delay} & 5 & \multicolumn{2}{c}{0.2966} & \multicolumn{2}{c}{0.0052} \\ 
        ~ & ~ & 10 & \multicolumn{2}{c}{0.2585} & \multicolumn{2}{c}{0.0050} \\ 
        ~ & ~ & 15 & \multicolumn{2}{c}{0.2346} & \multicolumn{2}{c}{0.0048} \\ \cdashline{2-7}
        ~ & \multirow{3}{0.5cm}{Pop III} & 5 & \multicolumn{2}{c}{0.3225} & \multicolumn{2}{c}{0.0052} \\ 
        ~ & ~ & 10 & \multicolumn{2}{c}{0.2833} & \multicolumn{2}{c}{0.0051} \\ 
        ~ & ~ & 15 & \multicolumn{2}{c}{0.2634} & \multicolumn{2}{c}{0.0049} \\
        \hline
        \textbf{Model} & \textbf{Source Type} & \textbf{Years} & \multicolumn{2}{c}{$\bm{\Delta H_0}$} & \multicolumn{2}{c}{$\bm{\Delta \Omega_m}$}\\
        \hline
        \multirow{9}{0.5cm}{PEDE} & \multirow{3}{0.5cm}{Delay} & 5 & \multicolumn{2}{c}{0.4033} & \multicolumn{2}{c}{0.0051} \\ 
        ~ & ~ & 10 & \multicolumn{2}{c}{0.3703} & \multicolumn{2}{c}{0.0049} \\ 
        ~ & ~ & 15 & \multicolumn{2}{c}{0.3284} & \multicolumn{2}{c}{0.0048} \\ \cdashline{2-7}
        ~ & \multirow{3}{0.5cm}{No Delay} & 5 & \multicolumn{2}{c}{0.3435} & \multicolumn{2}{c}{0.0049} \\ 
        ~ & ~ & 10 & \multicolumn{2}{c}{0.2834} & \multicolumn{2}{c}{0.0047} \\ 
        ~ & ~ & 15 & \multicolumn{2}{c}{0.2512} & \multicolumn{2}{c}{0.0045} \\ \cdashline{2-7}
        ~ & \multirow{3}{0.5cm}{Pop III} & 5 & \multicolumn{2}{c}{0.3823} & \multicolumn{2}{c}{0.005} \\ 
        ~ & ~ & 10 & \multicolumn{2}{c}{0.3213} & \multicolumn{2}{c}{0.0048} \\ 
        ~ & ~ & 15 & \multicolumn{2}{c}{0.2952} & \multicolumn{2}{c}{0.0047} \\ 
        \hline
        \textbf{Model} & \textbf{Source Type} & \textbf{Years} & \multicolumn{2}{c}{$\bm{\Delta H_0}$} & \multicolumn{2}{c}{$\bm{\Delta M}$}\\
        \hline
        \multirow{9}{0.5cm}{VM} & \multirow{3}{0.5cm}{Delay} & 5 & \multicolumn{2}{c}{0.5118} & \multicolumn{2}{c}{0.0020} \\ 
        ~ & ~ & 10 & \multicolumn{2}{c}{0.4462} & \multicolumn{2}{c}{0.0018} \\ 
        ~ & ~ & 15 & \multicolumn{2}{c}{0.4194} & \multicolumn{2}{c}{0.0017} \\ \cdashline{2-7}
        ~ & \multirow{3}{0.5cm}{No Delay} & 5 & \multicolumn{2}{c}{0.3955} &  \multicolumn{2}{c}{0.0018} \\ 
        ~ & ~ & 10 & \multicolumn{2}{c}{0.3118} & \multicolumn{2}{c}{0.0016} \\ 
        ~ & ~ & 15 & \multicolumn{2}{c}{0.2720} & \multicolumn{2}{c}{0.0014} \\ \cdashline{2-7}
        ~ & \multirow{3}{0.5cm}{Pop III} & 5 & \multicolumn{2}{c}{0.4623} & \multicolumn{2}{c}{0.0019} \\ 
        ~ & ~ & 10 & \multicolumn{2}{c}{0.3767} &\multicolumn{2}{c}{0.0017} \\ 
        ~ & ~ & 15 & \multicolumn{2}{c}{0.3381} & \multicolumn{2}{c}{0.0016} \\
        \hline
        \textbf{Model} & \textbf{Source Type} & \textbf{Years} & $\bm{\Delta H_0}$ & \multicolumn{2}{c}{$\bm{\Delta \Omega_m}$} & $\bm{\Delta M}$\\
        \hline
        \multirow{9}{0.5cm}{VM-VEV} & \multirow{3}{0.5cm}{Delay} & 5 & 0.2961 & \multicolumn{2}{c}{0.0039} & 0.0012 \\ 
        ~ & ~ & 10 & 0.2824 & \multicolumn{2}{c}{0.0038} & 0.0012 \\ 
        ~ & ~ & 15 & 0.2678 & \multicolumn{2}{c}{0.0037} & 0.0012 \\ \cdashline{2-7}
        ~ & \multirow{3}{0.5cm}{No Delay} & 5 & 0.2659 & \multicolumn{2}{c}{0.0037} & 0.0012 \\ 
        & ~ & 10 & 0.2367 & \multicolumn{2}{c}{0.0035} & 0.0011 \\ 
        ~ & ~ & 15 & 0.2170 & \multicolumn{2}{c}{0.0034} & 0.0011 \\ \cdashline{2-7}
        ~ & \multirow{3}{0.5cm}{Pop III} & 5 & 0.2844 & \multicolumn{2}{c}{0.0038} & 0.0012 \\ 
        ~ & ~ & 10 & 0.2601 & \multicolumn{2}{c}{0.0036} & 0.0011 \\ 
        ~ & ~ & 15 & 0.2463 & \multicolumn{2}{c}{0.0035} & 0.0011 \\  
        \hline
        \textbf{Model} & \textbf{Source Type} & \textbf{Years} & $\bm{\Delta H_0}$ & $\bm{\Delta \Omega_m}$ & $\bm{\Delta w_0}$ & $\bm{\Delta w_a}$ \\
        \hline
        \multirow{9}{0.5cm}{CPL} & \multirow{3}{0.5cm}{Delay} & 5 & 0.6903 & 0.0077 & 0.0661 & 0.2710 \\ 
        ~ & ~ & 10 & 0.6571 & 0.0075 & 0.0632 & 0.2614 \\ 
        ~ & ~ & 15 & 0.6540 & 0.0074 & 0.0622 & 0.2566 \\ \cdashline{2-7}
        ~ & \multirow{3}{0.5cm}{No Delay} & 5 & 0.6360 & 0.0076 & 0.0618 & 0.2590 \\ 
        ~ & ~ & 10 & 0.5866 & 0.0074 & 0.0605 & 0.2455 \\ 
        ~ & ~ & 15 & 0.5586 & 0.0073 & 0.0591 & 0.2356 \\ \cdashline{2-7}
        ~ & \multirow{3}{0.5cm}{Pop III} & 5 & 0.6916 & 0.0077 & 0.0643 & 0.2672 \\
        ~ & ~ & 10 & 0.6292 & 0.0075 & 0.0622 & 0.2549 \\ 
        ~ & ~ & 15 & 0.5935 & 0.0074 & 0.0608 & 0.2444 \\
        \hline
        \textbf{Model} & \textbf{Source Type} & \textbf{Years} & $\bm{\Delta H_0}$ & $\bm{\Delta \Omega_m}$ & $\bm{\Delta w_0}$ & $\bm{\Delta w_a}$ \\
        \hline
        \multirow{9}{0.5cm}{JBP} & \multirow{3}{0.5cm}{Delay} & 5 & 0.7335 & 0.0075 & 0.0929 & 0.6157 \\ 
        ~ & ~ & 10 & 0.7052 & 0.0074 & 0.0886 & 0.5872 \\ 
        ~ & ~ & 15 & 0.6911 & 0.0073 & 0.0858 & 0.5666 \\ \cdashline{2-7}
        ~ & \multirow{3}{0.5cm}{No Delay} & 5 & 0.6682 & 0.0074 & 0.086 & 0.5708 \\ 
        ~ & ~ & 10 & 0.6208 & 0.0073 & 0.0828 & 0.5265 \\ 
        ~ & ~ & 15 & 0.5679 & 0.0072 & 0.0807 & 0.4939 \\ \cdashline{2-7}
        ~ & \multirow{3}{0.5cm}{Pop III} & 5 & 0.6921 & 0.0075 & 0.0873 & 0.5902 \\ 
        ~ & ~ & 10 & 0.6688 & 0.0074 & 0.0855 & 0.5647 \\ 
        ~ & ~ & 15 & 0.6273 & 0.0073 & 0.0842 & 0.5355 \\ 
        \hline
    \end{tabular}
    }
\caption{$1\sigma$ errors from Fisher analysis of the simulated standard siren catalogs for each cosmological model, MBHB source type, and eLISA mission duration. The errors $\Delta H_0$ are measured in units of km s\textsuperscript{-1} Mpc\textsuperscript{-1}.}
    \label{Tab:Fisher}
\end{table}

\begin{figure}[!h]
    \centering
    \includegraphics[width=\textwidth,height=0.45\textheight,keepaspectratio]{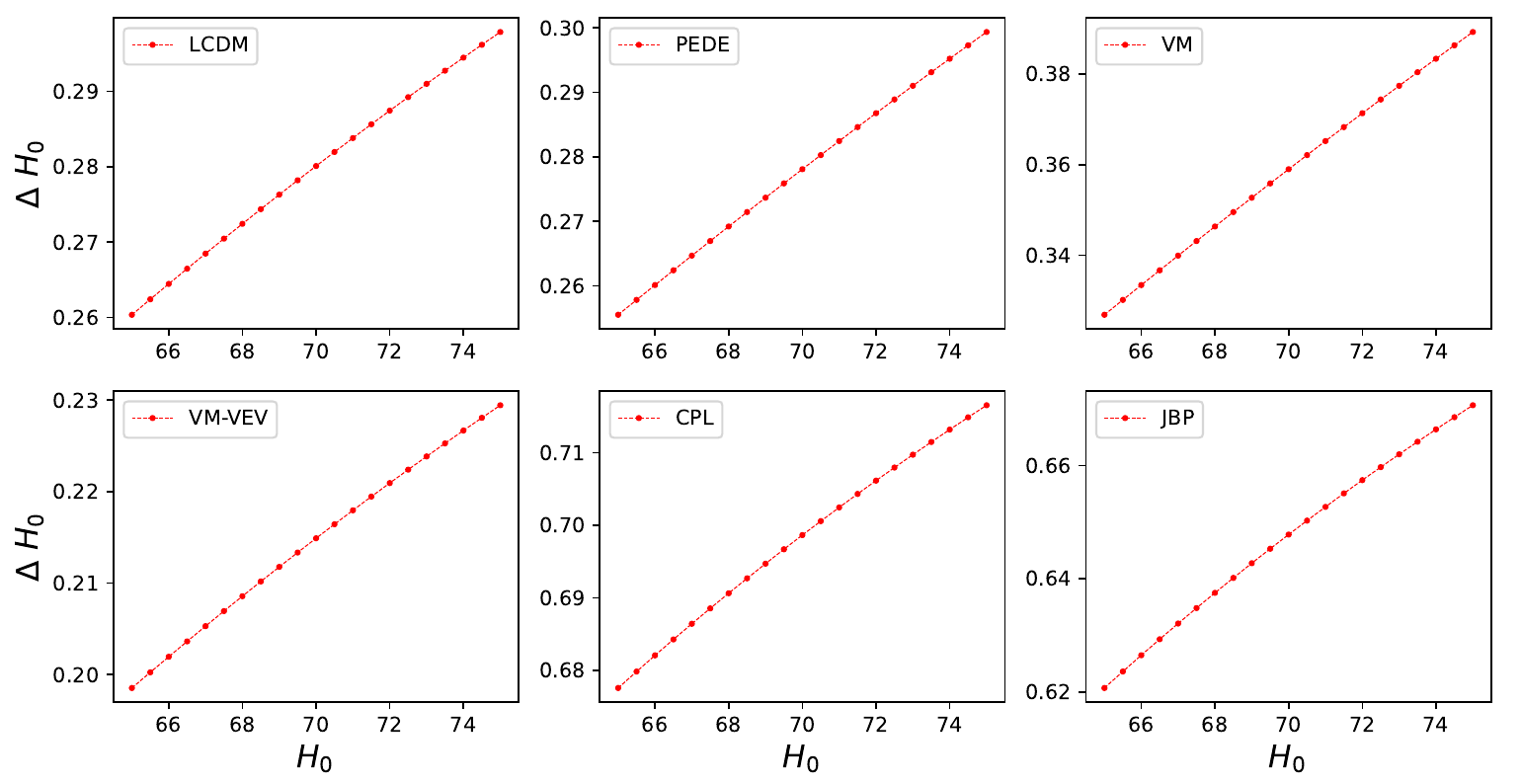}
    \caption{Dependence of errors estimated by Fisher analysis on the mean value of $H_0$ (km s\textsuperscript{-1} Mpc\textsuperscript{-1}) for source type No Delay and eLISA mission duration of 10 years.}
    \label{fig:fishererrvsmean}
\end{figure}

Considering the three source types mentioned in Sec. \ref{sec:catalogue}, the Fisher analysis results for the cosmological models under consideration, corresponding to eLISA operating durations of 5, 10, and 15 years, are listed in Table \ref{Tab:Fisher}. For an unbiased analysis, we have run the Fisher prescription on all 500 catalogs for each model, source population, and mission duration. The priors have been obtained directly from the constraints listed in Table \ref{constraints}. The $1\sigma$ error value quoted under each category in the tables reflects an average across individual error forecasts from all the catalogs under that category, which are found to be mostly consistent. The contour plots for various MBHB source types corresponding to a typical eLISA mission of 10 years duration are shown in Figs. \ref{fig:fisherdelay}, \ref{fig:fishernodelay}, and \ref{fig:fisherpopIII} (plotted using the Python library of \textit{CosmicFish} \cite{cosmicfish1,cosmicfish2}) for a representative catalog. Our currently adopted methodology to obtain fair estimates of the one-parameter errors and the two-parameter contours is inspired by \cite{eLISA3}.

To complete our exposition of the Fisher methodology, we show in Fig. \ref{fig:fishererrvsmean} the variation of the predicted errors ($\Delta H_0$) across a range of plausible mean $H_0$ values detectable in reality by eLISA, corresponding to the source type ``No Delay'' and mission duration of 10 years. As visible in the graph, the behavior shows slight deviation from linearity. Moreover, the $H_0$ error forecasts are not too dependent (being mostly at the sub-percent level) on the choice of the mean values within the range of our interest.


\subsubsection{\label{sec:GWMCMC}Markov Chain Monte Carlo}

As the second approach, we perform a Markov Chain Monte Carlo analysis using our generated catalogs to constrain cosmological parameters, using the Python package \textit{emcee} \cite{Foreman_Mackey_2013}. For every model, mission duration, and source population, we have run MCMC on 100 catalogs for an unbiased analysis. We compute the median for the constraints obtained and show the results for a representative catalog which lies closest to this joint estimate.

\begin{table*}[!h]
\centering
    \resizebox{\textwidth}{!}{\renewcommand{\arraystretch}{1.4} \setlength{\tabcolsep}{15.5 pt} \centering 
    \begin{tabular}{c c c c c c c c}
        \hline\hline
\textbf{Source Type}  &  \textbf{Parameter} & $\bm{\Lambda}$\textbf{CDM} & \textbf{CPL} & \textbf{JBP} & \textbf{PEDE} & \textbf{VM} & \textbf{VM-VEV} \\ \hline
        \multicolumn{8}{c}{\textbf{5 Years}}\\ \hline
        &  $H_0$ & $67.256^{+0.782}_{-0.873}$ & $67.395^{+2.049}_{-2.042}$ & $68.892^{+1.433}_{-1.413}$ & $70.488^{+0.872}_{-1.069}$ & $73.808^{+0.643}_{-0.642}$ & $71.233^{+0.710}_{-0.784}$  \\
        &  $\Omega_{m0}$ & $0.313^{+0.019}_{-0.017}$ & $0.310^{+0.025}_{-0.025}$ & $0.302^{+0.024}_{-0.026}$ & $0.298^{+0.018}_{-0.016}$ & $-$ & $0.281^{+0.014}_{-0.013}$  \\
Delay   &  $w_0$ & $-$ & $-0.983^{+0.219}_{-0.215}$ & $-1.150^{+0.206}_{-0.204}$ & $-$ & $-$ & $-$  \\
        &  $w_a$ & $-$ & $-0.898^{+0.499}_{-0.471}$ & $-0.294^{+0.842}_{-0.811}$ & $-$ & $-$ & $-$  \\
        &  $M$ & $-$ & $-$ & $-$ & $-$ & $ 0.906^{+0.006}_{-0.006}$ & $0.853^{+0.014}_{-0.015}$  \\
        \hline
        &  $H_0$ & $67.868^{+0.334}_{-0.345}$ & $68.097^{+1.713}_{-1.625}$ & $69.120^{+0.905}_{-0.871}$ & $71.488^{+0.419}_{-0.419}$ & $73.873^{+0.419}_{-0.413}$ & $72.681^{+0.431}_{-0.460}$  \\
        &  $\Omega_{m0}$ & $0.309^{+0.010}_{-0.009}$ & $0.312^{+0.025}_{-0.030}$ & $0.301^{+0.014}_{-0.015}$ & $0.290^{+0.009}_{-0.009}$ & $-$ & $0.274^{+0.010}_{-0.010}$  \\
No Delay & $w_0$ & $-$ & $-0.956^{+0.186}_{-0.202}$ & $-1.098^{+0.131}_{-0.137}$ & $-$ & $-$ & $-$  \\
        &  $w_a$ & $-$ & $-0.712^{+0.753}_{-0.702}$ & $-0.526^{+0.715}_{-0.664}$ & $-$ & $-$ & $-$  \\
        &  $M$ & $-$ & $-$ & $-$ & $-$ & $0.916^{+0.004}_{-0.004}$ & $0.877^{+0.009}_{-0.013}$  \\
        \hline
        &  $H_0$ & $67.928^{+0.573}_{-0.584}$ & $69.738^{+1.891}_{-1.792}$ & $69.307^{+1.010}_{-0.988}$ & $70.534^{+0.558}_{-0.557}$ & $73.828^{+0.568}_{-0.564}$ & $72.211^{+0.523}_{-0.566}$  \\
        &  $\Omega_{m0}$ & $ 0.305^{+0.015}_{-0.014}$ & $0.292^{+0.028}_{-0.033}$ & $0.294^{+0.014}_{-0.015}$ & $0.299^{+0.013}_{-0.012}$ & $-$ & $0.273^{+0.011}_{-0.010}$  \\
Pop III     & $w_0$ & $-$ & $-1.065^{+0.240}_{-0.255}$ & $-1.124^{+0.128}_{-0.129}$ & $-$ & $-$ & $-$  \\
        &  $w_a$ & $-$ & $-0.715^{+0.879}_{-0.839}$ & $-0.342^{+0.623}_{-0.583}$ & $-$ & $-$ & $-$  \\
        &  $M$ & $-$ & $-$ & $-$ & $-$ & $0.914^{+0.005}_{-0.005}$ & $0.869^{+0.010}_{-0.013}$  \\
        \hline
        \multicolumn{8}{c}{\textbf{10 Years}}\\ \hline
        &  $H_0$ & $67.348^{+0.391}_{-0.395}$ & $68.552^{+1.089}_{-1.059}$ & $69.284^{+1.034}_{-1.029}$ & $71.061^{+0.454}_{-0.461}$ & $73.814^{+0.521}_{-0.524}$ & $72.115^{+0.469}_{-0.479}$  \\
        &  $\Omega_{m0}$ & $0.314^{+0.009}_{-0.009}$ & $0.295^{+0.015}_{-0.018}$ & $0.299^{+0.014}_{-0.015}$ & $0.290^{+0.010}_{-0.009}$ & $ $ & $0.269^{+0.008}_{-0.008}$  \\
Delay   &  $w_0$ & $-$ & $-0.909^{+0.126}_{-0.130}$ & $-1.160^{+0.131}_{-0.133}$ & $-$ & $-$ & $-$  \\
        &  $w_a$ & $-$ & $-0.659^{+0.512}_{-0.514}$ & $-0.287^{+0.629}_{-0.603}$ & $-$ & $-$ & $-$  \\
        &  $M$ & $-$ & $-$ & $-$ & $-$ & $0.906^{+0.005}_{-0.005}$ & $0.869^{+0.011}_{-0.013}$  \\
        \hline
        &  $H_0$ & $68.181^{+0.271}_{-0.270}$ & $68.736^{+0.847}_{-0.824}$ & $69.166^{+0.718}_{-0.737}$ & $71.384^{+0.286}_{-0.286}$ & $73.849^{+0.357}_{-0.356}$ & $72.697^{+0.336}_{-0.361}$  \\
        &  $\Omega_{m0}$ & $0.301^{+0.008}_{-0.008}$ & $0.294^{+0.021}_{-0.024}$ & $0.293^{+0.013}_{-0.014}$ & $0.285^{+0.007}_{-0.007}$ & $-$ & $0.274^{+0.008}_{-0.008}$  \\
No Delay &  $w_0$ & $-$ & $-0.947^{+0.101}_{-0.107}$ & $-1.067^{+0.112}_{-0.113}$ & $-$ & $-$ & $-$  \\
        &  $w_a$ & $-$ & $-0.208^{+0.605}_{-0.665}$ & $0.054^{+0.763}_{-0.760}$ & $-$ & $-$ & $-$  \\
        &  $M$ & $-$ & $-$ & $-$ & $-$ & $0.917^{+0.003}_{-0.003}$ & $0.884^{+0.005}_{-0.007}$  \\
        \hline
        &  $H_0$ & $67.727^{+0.356}_{-0.353}$ & $68.735^{+0.927}_{-0.898}$ & $69.733^{+0.767}_{-0.760}$ & $71.613^{+0.333}_{-0.330}$ & $73.821^{+0.409}_{-0.407}$ & $72.779^{+0.413}_{-0.416}$  \\
        &  $\Omega_{m0}$ & $0.309^{+0.009}_{-0.009}$ & $0.304^{+0.015}_{-0.016}$ & $0.309^{+0.010}_{-0.010}$ & $0.277^{+0.007}_{-0.007}$ & $-$ & $0.265^{+0.011}_{-0.010}$  \\
Pop III  &  $w_0$ & $-$ & $-0.967^{+0.109}_{-0.116}$ & $-1.195^{+0.114}_{-0.112}$ & $-$ & $-$ & $-$  \\
        &  $w_a$ & $-$ & $-0.531^{+0.532}_{-0.522}$ & $-0.067^{+0.689}_{-0.680}$ & $-$ & $-$ & $-$  \\
        &  $M$ & $-$ & $-$ & $-$ & $-$ & $0.913^{+0.004}_{-0.004}$ & $0.883^{+0.008}_{-0.010}$  \\
        \hline
        \multicolumn{8}{c}{\textbf{15 Years}}\\ \hline
        &  $H_0$ & $67.910^{+0.305}_{-0.314}$ & $68.902^{+0.922}_{-0.865}$ & $69.153^{+0.823}_{-0.816}$ & $ 70.968^{+0.374}_{-0.384}$ & $ 73.781^{+0.405}_{-0.402}$ & $72.867^{+0.360}_{-0.390}$  \\
        &  $\Omega_{m0}$ & $0.306^{+0.007}_{-0.007}$ & $0.299^{+0.012}_{-0.013}$ & $0.293^{+0.009}_{-0.009}$ & $0.290^{+0.007}_{-0.007}$ & $-$ & $0.272^{+0.007}_{-0.007}$  \\
Delay   &  $w_0$ & $-$ & $-0.987^{+0.099}_{-0.102}$ & $-1.108^{+0.098}_{-0.096}$ & $-$ & $-$ & $-$  \\
        &  $w_a$ & $-$ & $-0.597^{+0.382}_{-0.379}$ & $-0.463^{+0.471}_{-0.453}$ & $-$ & $-$ & $-$  \\
        &  $M$ & $-$ & $-$ & $-$ & $-$ & $0.909^{+0.004}_{-0.004}$ & $0.882^{+0.007}_{-0.010}$  \\
        \hline
        &  $H_0$ & $68.016^{+0.239}_{-0.240}$ & $68.864^{+0.628}_{-0.601}$ & $68.820^{+0.643}_{-0.642}$ & $71.259^{+0.252}_{-0.248}$ & $73.816^{+0.305}_{-0.309}$ & $73.489^{+0.309}_{-0.307}$  \\
        &  $\Omega_{m0}$ & $0.301^{+0.006}_{-0.006}$ & $0.301^{+0.017}_{-0.022}$ & $0.302^{+0.010}_{-0.011}$ & $0.290^{+0.006}_{-0.006}$ & $-$ & $0.275^{+0.012}_{-0.012}$  \\
No Delay &  $w_0$ & $-$ & $-0.886^{+0.089}_{-0.084}$ & $-1.054^{+0.109}_{-0.108}$ & $-$ & $-$ & $-$  \\
        &  $w_a$ & $-$ & $-0.263^{+0.561}_{-0.612}$ & $-0.105^{+0.742}_{-0.731}$ & $-$ & $-$ & $-$  \\
        &  $M$ & $-$ & $-$ & $-$ & $-$ & $0.915^{+0.003}_{-0.003}$ & $0.898^{+0.004}_{-0.005}$  \\
        \hline
        &  $H_0$ & $67.929^{+0.285}_{-0.283}$ & $68.837^{+0.666}_{-0.655}$ & $68.832^{+0.697}_{-0.699}$ & $70.942^{+0.315}_{-0.308}$ & $73.670^{+0.357}_{-0.355}$ & $72.533^{+0.322}_{-0.325}$  \\
        &  $\Omega_{m0}$ & $0.308^{+0.007}_{-0.007}$ & $0.309^{+0.012}_{-0.015}$ & $0.306^{+0.013}_{-0.014}$ & $0.291^{+0.007}_{-0.006}$ & $-$ & $0.264^{+0.008}_{-0.008}$  \\
Pop III  &  $w_0$ & $-$ & $-0.913^{+0.088}_{-0.093}$ & $-1.029^{+0.114}_{-0.114}$ & $-$ & $-$ & $-$  \\
        &  $w_a$ & $-$ & $-0.559^{+0.514}_{-0.508}$ & $-0.340^{+0.818}_{-0.769}$ & $-$ & $-$ & $-$  \\
        &  $M$ & $-$ & $-$ & $-$ & $-$ & $0.912^{+0.003}_{-0.003}$ & $0.876^{+0.007}_{-0.009}$  \\
        \hline
        \hline
    \end{tabular}
    }
    \caption{Marginalized constraints on the parameters of the cosmological models considered in Sec. \ref{sec:models} for eLISA using MCMC.}
    \label{tab:gw_mcmc}
\end{table*}

\begin{figure}[!t]
    \centering
    \includegraphics[width=\textwidth,height=0.41\textheight,keepaspectratio]{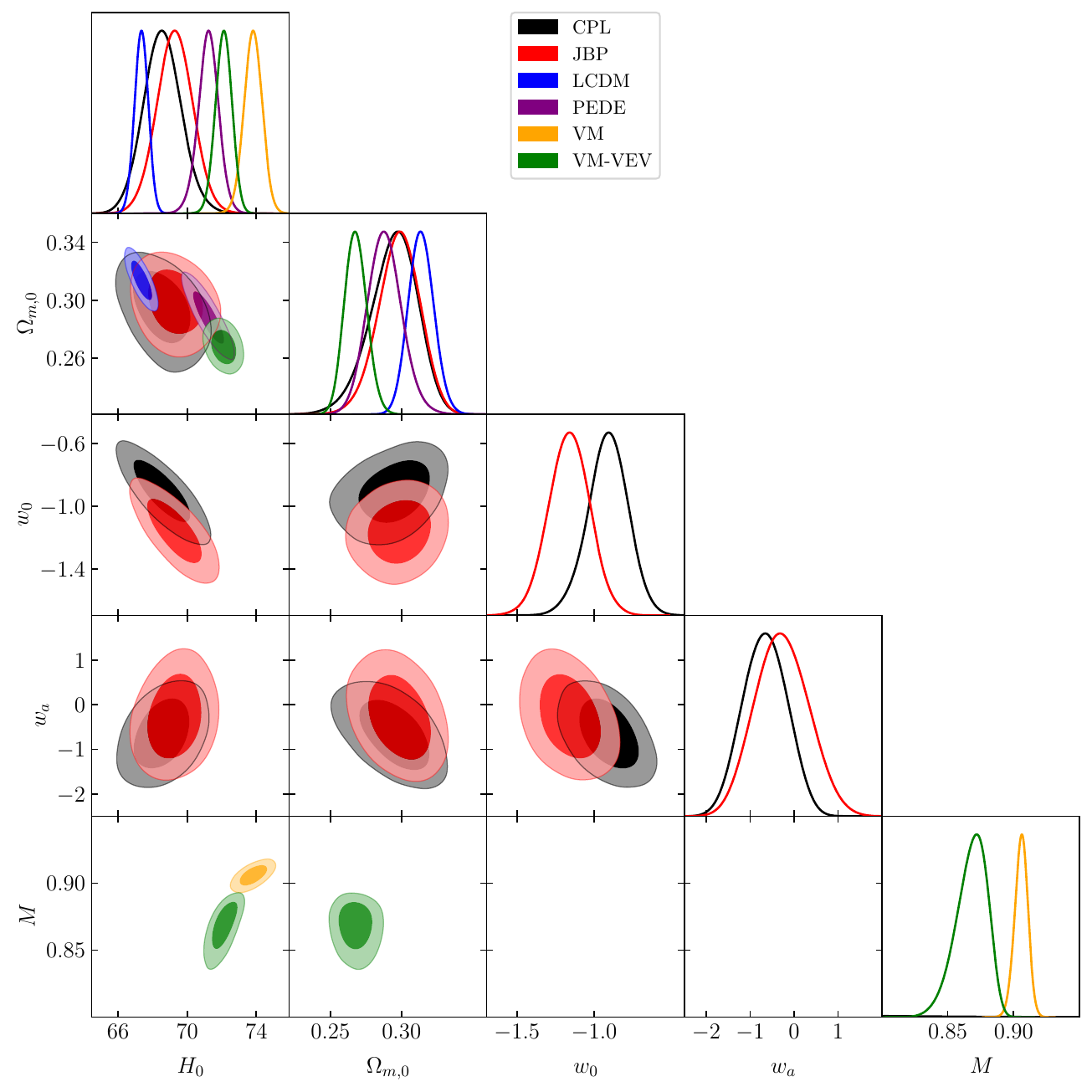}
    \caption{Markov Chain Monte Carlo contours with Delay source type for 10 years eLISA mission duration.}
    \label{mcmcdelay}
\end{figure}

\begin{figure}[!t]
    \centering
    \includegraphics[width=\textwidth,height=0.41\textheight,keepaspectratio]{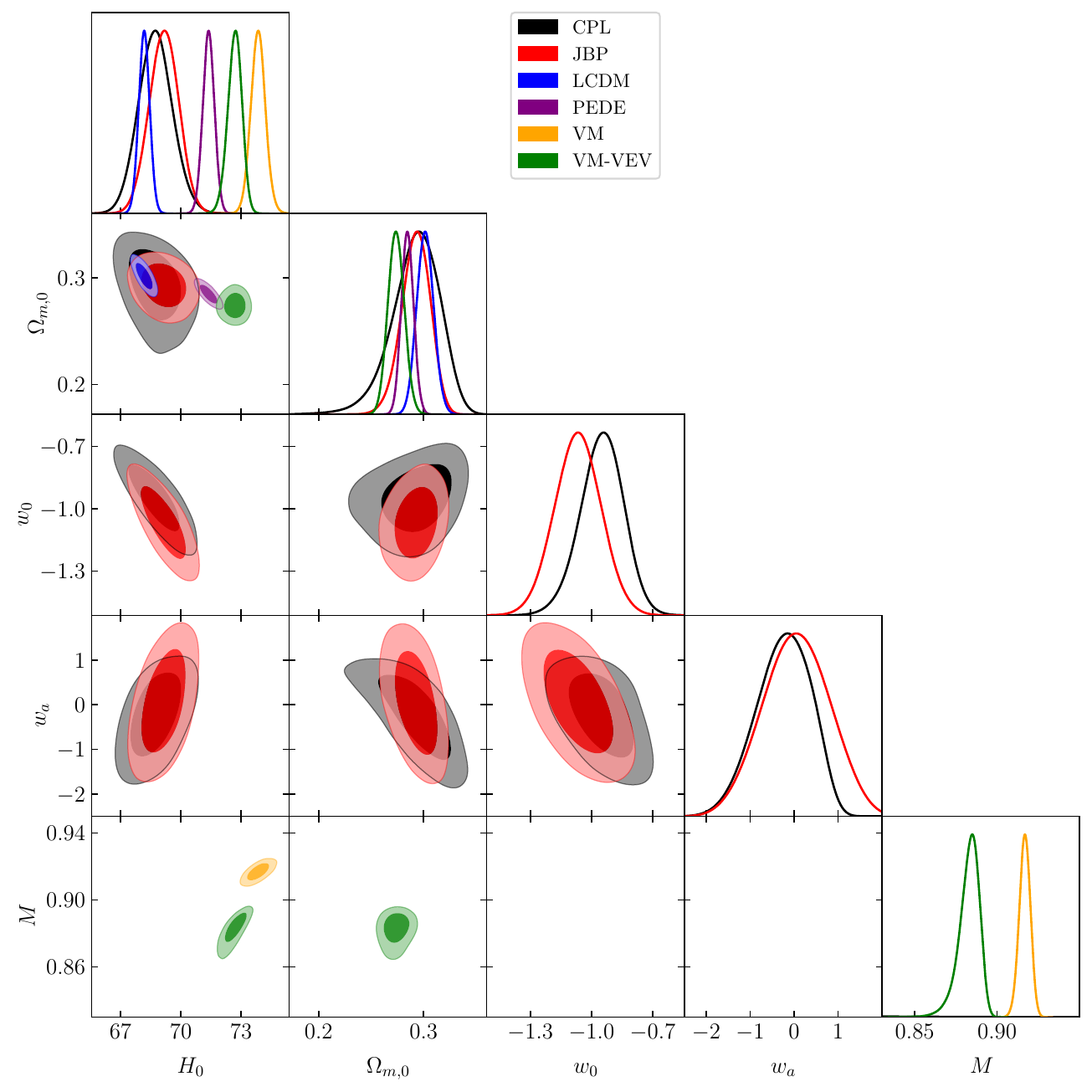}
    \caption{Markov Chain Monte Carlo contours with No Delay source type for 10 years eLISA mission duration.}
    \label{mcmcnodelay}
\end{figure}

\begin{figure}[!t]
    \centering
    \includegraphics[width=\textwidth,height=0.41\textheight,keepaspectratio]{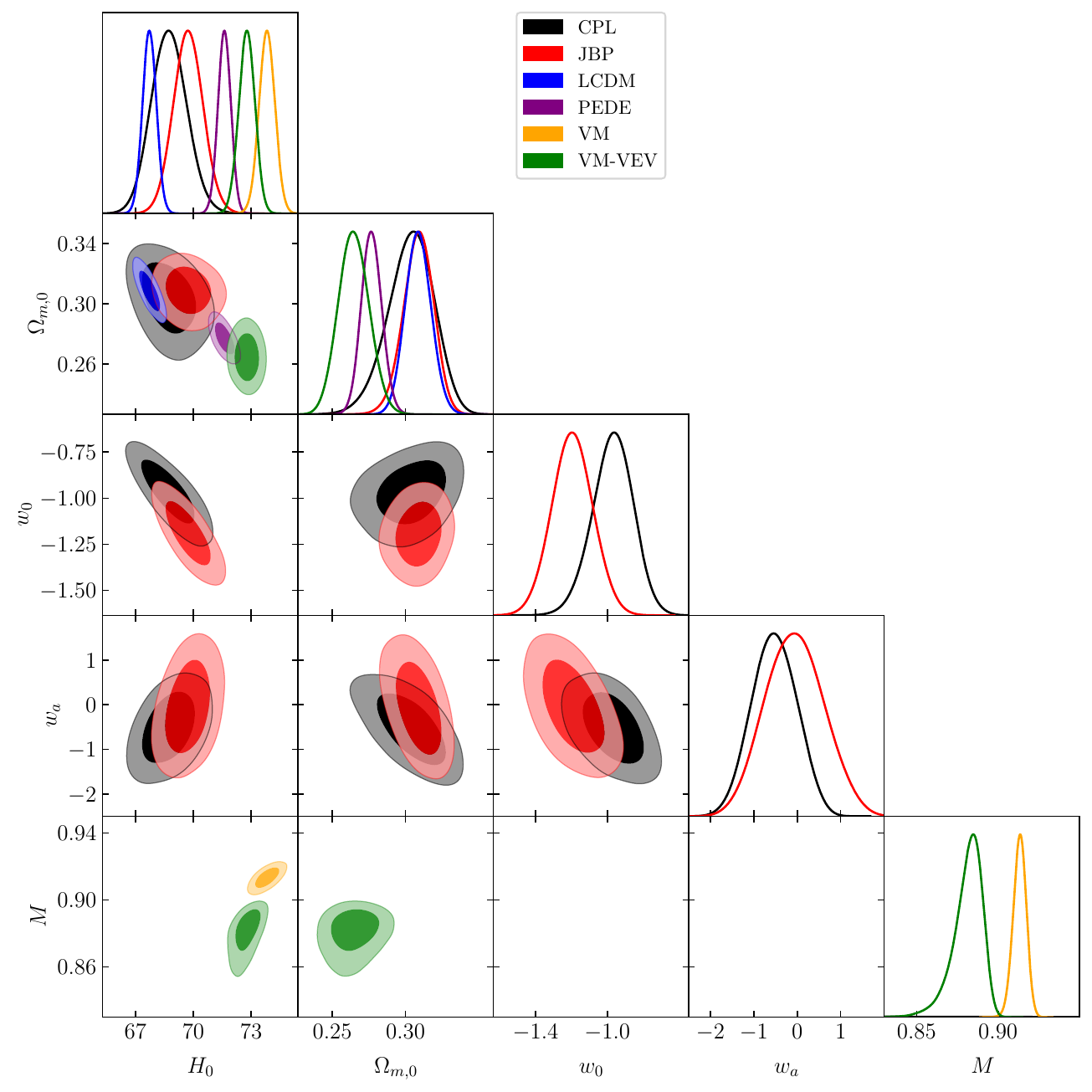}
    \caption{Markov Chain Monte Carlo contours with Pop III source type for 10 years eLISA mission duration.}
    \label{mcmcpopIII}
\end{figure}

To estimate the cosmological parameter values from observational data, the chi-squared statistic is employed here, which is given as
\begin{equation}
\chi^2=\sum_i^N \frac{\left[d_{L_i}-d_L\left(z_i,\{\theta\}\right)\right]^2}{\sigma_i^2}\:\:,
\end{equation}
where we have the data in the catalog $(z_i,d_{L_i})$ for $i=\{1,...,N\}$, with $\sigma_i$ being the corresponding noise of each measurement, and $d_L\left(z_i,\{\theta\}\right)$ being the theoretical function to describe the dataset with the set of parameters $\{\theta\}$. Since the observations are at distinct redshifts, we assume no correlation between them. We assume priors on the parameters as given in Table \ref{tab:priors}. The results are summarized in Table \ref{tab:gw_mcmc}. The contour plots for 10 years of each mission duration are shown in Figs. \ref{mcmcdelay}, \ref{mcmcnodelay} and \ref{mcmcpopIII}, using \textit{GetDist} \cite{Lewis:2019xzd}.


\subsection{Non-parametric method}

\subsubsection{\label{sec:GP}Machine Learning with Gaussian Processes}

While the previous methods fall under the category of parametric approaches, there can be alternative, non-parametric approaches to the problem under consideration. To this end, we use the so-called Gaussian Processes (GP) \cite{Holsclaw_2011,Seikel:2012uu,Shafieloo_2012}, a non-parametric machine learning tool, to infer the present value of the Hubble parameter from the generated eLISA catalogs. Gaussian Processes, being distributions over functions, are essentially generalizations of Gaussian distributions over variables. For our purpose, given a set of labeled training data, GP can be used to reconstruct the underlying most probable continuous function describing that dataset along with the associated 1$\sigma$ uncertainties, without assuming any parametric cosmological model. GP has found widespread application in cosmological reconstructions (see \cite{Mukherjee:2022lkt} and the references therein). For a general overview, one can refer to the Gaussian Process website\footnote{\url{http://www.gaussianprocess.org}}. 

\begin{figure*}
    \centering
    \begin{subfigure}{\textwidth}
        \centering        \includegraphics[width=0.95\textwidth]{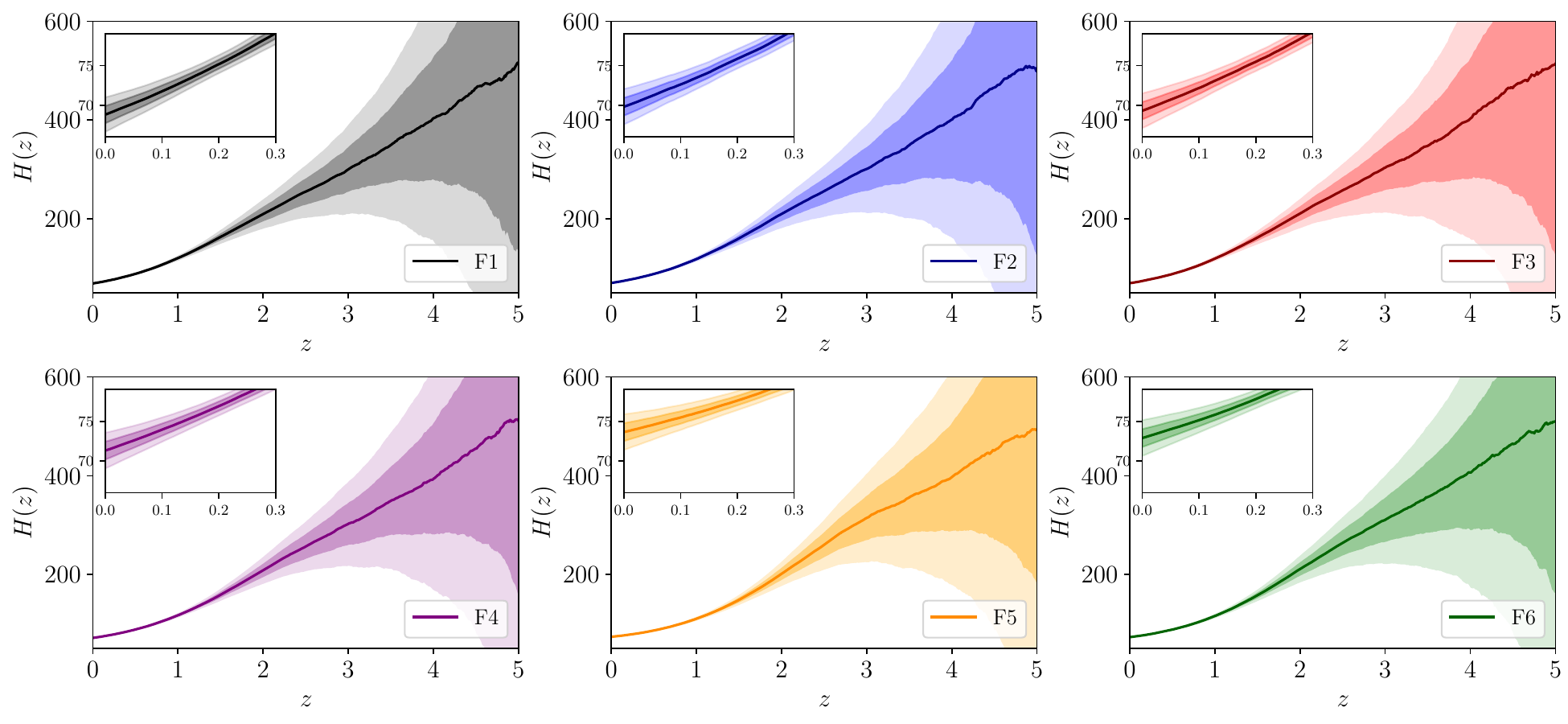}
        \caption{Pop III}
    \end{subfigure}
    \begin{subfigure}{\textwidth}
        \centering        \includegraphics[width=0.95\textwidth]{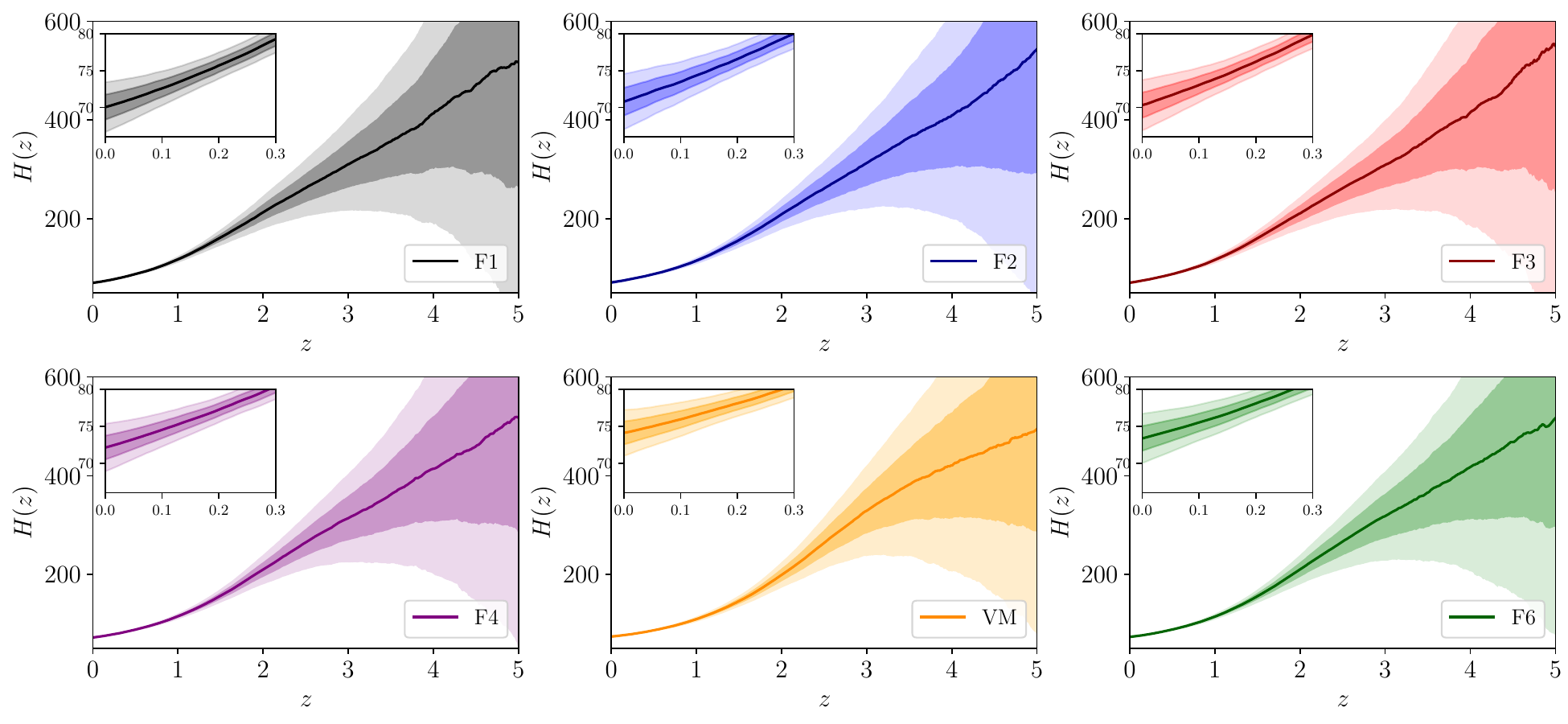}
        \caption{Delay}
    \end{subfigure}
    \begin{subfigure}{\textwidth}
        \centering        \includegraphics[width=0.95\textwidth]{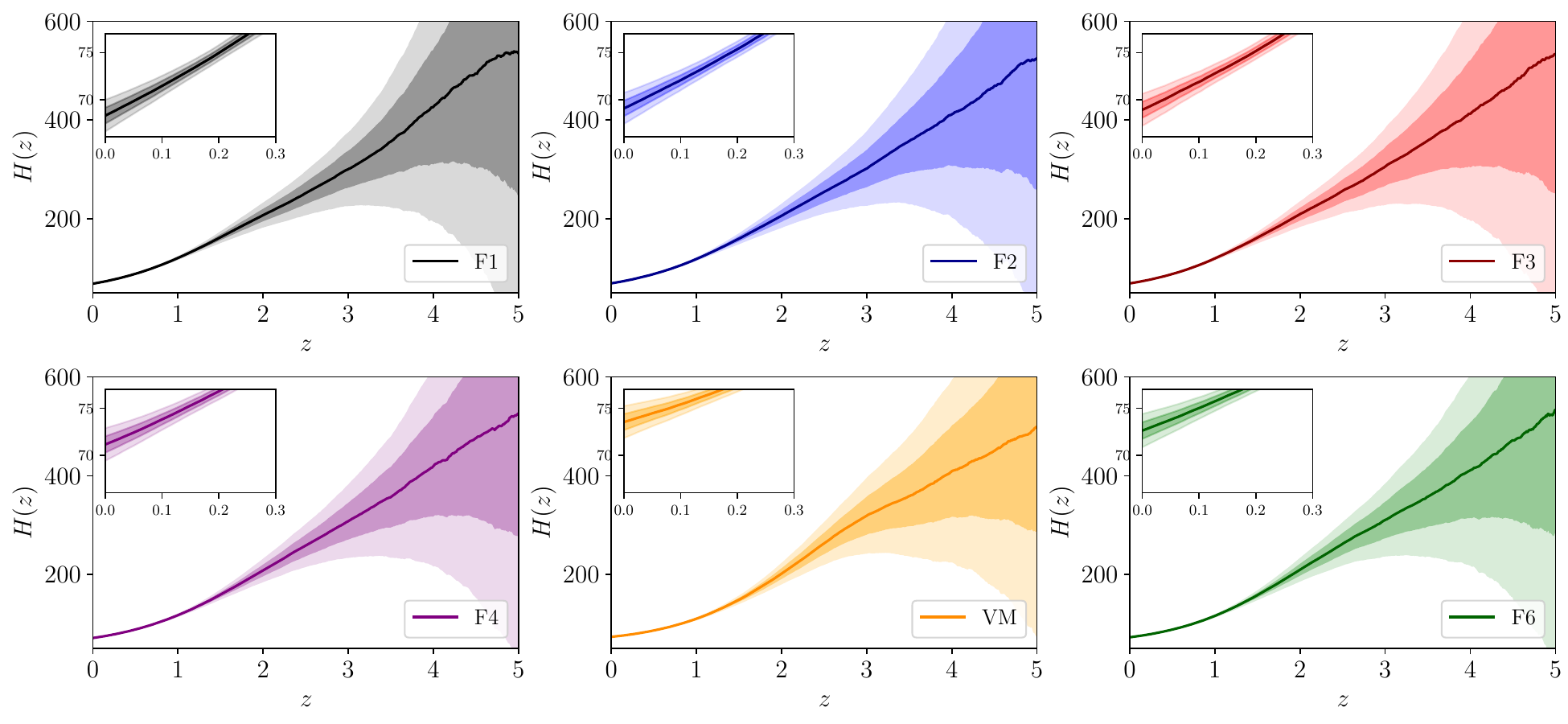}
        \caption{No Delay}
    \end{subfigure}
    \caption{Plots for the reconstructed $H(z)$ (km s\textsuperscript{-1} Mpc\textsuperscript{-1}) with redshift $z$ using GP for 10 year eLISA mission. The solid line represents the mean curve. The associated 1$\sigma$ and 2$\sigma$ confidence regions are shown in lighter shades.} \label{fig:gp_Hz}	
\end{figure*}

In a spatially flat Universe described by the Friedmann–Robertson–Lemaître–Walker (FRLW) metric, the Hubble parameter $H(z)$ can be expressed as
\begin{equation}
    H(z) = \frac{c (1+z)^2}{{d_L}^\prime (1+z) -d_L}\:\:,
\end{equation}
where $d_L$ is the luminosity distance function and ${d_L}^\prime$ is its first order derivative with respect to redshift ($z$).

Using the sets of 500 mock catalogs mentioned in Sec. \ref{sec:catalogue}, we utilize GP to reconstruct $d_L$, ${d_L}^\prime$ and the covariance between $d_L$ and ${d_L}^\prime$ (\textit{i.e.}, $\text{Cov}[d_L, {d_L}^\prime]$). In order to emphasize the overall fiducial-dependence as opposed to explicit model-dependence, we have renamed the catalog sets F1, F2, F3, F4, F5, and F6 respectively (for a given source type and mission duration) when used in the context of GP. Throughout this work, we assume a zero mean function and the Mat\'{e}rn 9/2 covariance function, as suggested in \cite{Seikel:2013fda} to characterize the GP. We have followed the marginalized approach similar to \cite{Hwang:2022hla, Banerjee:2023rvg} to prevent any overfitting issues \cite{OColgain:2021pyh}. 

\begin{table*}[!t] 
    \resizebox{\textwidth}{!}{\renewcommand{\arraystretch}{2.0} \setlength{\tabcolsep}{5 pt} \centering  
    \begin{tabular}{c c c c c c c c }
        \hline\hline
        \textbf{Years} & \textbf{Source Type} & \textbf{F1} & \textbf{F2} & \textbf{F3} & \textbf{F4} & \textbf{F5} & \textbf{F6}\\ \hline	
        & Delay  & $73.935 \pm 4.213$ & $74.416 \pm 4.386$ & $74.559 \pm 4.379$ & $75.494 \pm 4.356$ & $75.320 \pm 4.276$ & $75.921 \pm 4.598$ \\
        5 & No Delay & $68.871 \pm 1.179$ & $69.740 \pm 1.182$ & $69.487 \pm 1.185$ & $71.459 \pm 1.175$ & $73.868 \pm 1.139$ & $72.929 \pm 1.153$ \\
        & Pop III & $ 70.047 \pm 1.945 $ & $ 70.610 \pm 1.662 $ & $70.268 \pm 1.667$ & $72.307 \pm 1.700$ & $74.333 \pm 1.743$ & $73.577 \pm 1.830$ \\
        \hline  
        & Delay  & $70.054 \pm 1.697$ & $70.795 \pm 1.884$ & $70.335 \pm 1.720$ & $72.126 \pm 1.615$ & $74.120 \pm 1.561$ & $73.374 \pm 1.680$ \\
        10 & No Delay & $68.281 \pm 0.846$ & $69.051 \pm 0.829$ & $68.901 \pm 0.855$ & $71.145 \pm 0.871$ & $73.537 \pm 0.856$ & $72.625 \pm 0.889$ \\
        & Pop III & $68.827 \pm 1.092$ & $69.813 \pm 1.122$ & $69.311 \pm 1.099$ & $71.318 \pm 1.124$ & $73.638 \pm 1.128$ & $72.889 \pm 1.130$ \\
        \hline
        & Delay  & $69.178 \pm 1.242$ & $70.075 \pm 1.361$ & $69.650 \pm 1.233$ & $71.531 \pm 1.205$ & $73.601 \pm 1.237$ & $72.858 \pm 1.195$ \\
        15 & No Delay & $68.142 \pm 0.707$ & $68.915 \pm 0.701$ & $68.794 \pm 0.715$ & $71.012 \pm 0.710$ & $73.478 \pm 0.729$ & $72.595 \pm 0.733$ \\
        & Pop III & $68.367 \pm 0.941$ & $69.241 \pm 0.915$ & $68.857 \pm 0.907$ & $71.128 \pm 0.905$ & $73.665 \pm 0.899$ & $72.639 \pm 0.902$ \\
        \hline \hline
    \end{tabular}
    }
    \caption{Table showing the reconstructed values of $H_0$ (in units of km s\textsuperscript{-1} Mpc\textsuperscript{-1}) using GP.} \label{tab:gp_H0}
\end{table*}

With these reconstructed $d_L (z)$ and ${d_L}^\prime (z)$, we can derive the evolution of $H(z)$ as a function of redshift and infer $H_0$ directly. The averaged $H_0$ results for each set of these 500 realizations are presented in Table \ref{tab:gp_H0}. We have also shown the reconstructed $H(z)$ functions for 10 years of eLISA operation in Fig. \ref{fig:gp_Hz} for an exhaustive presentation of the methodology adopted.


\section{\label{sec:analysis}Analysis and discussions}

In this section, we analyze the results obtained in the previous section based on the three different approaches. We also make a comparison among them and discuss the possible pros and cons of each individual approach. 

\begin{table*}[!ht] 
    \resizebox{\textwidth}{!}{\renewcommand{\arraystretch}{1.6} \setlength{\tabcolsep}{10 pt} \centering  
    \begin{tabular}{c | c c c c c c }
        \hline\hline
        \textbf{Model/Fiducial} & $\bm{\Lambda}$\textbf{CDM/F1} & \textbf{CPL/F2} & \textbf{JBP/F3} & \textbf{PEDE/F4} & \textbf{VM/F5} & \textbf{VM-VEV/F6}\\ \hline	
        Current Datasets & $4.98\sigma$ & $3.69\sigma$& $3.78\sigma$ & $1.79\sigma$ & $0.74\sigma$ & $0.04\sigma$ \\ \cline{1-7}
        Fisher Forecasting & $5.20\sigma$ & $4.15\sigma$ & $4.11\sigma$ & $1.91\sigma$ & $0.84\sigma$ & $0.04\sigma$ \\
        GW MCMC & $4.76\sigma$ & $3.40\sigma$ & $3.27\sigma$ & $1.78\sigma$ & $0.50\sigma$ & $0.55\sigma$ \\
        Gaussian Processes & $3.74\sigma$ & $3.19\sigma$ & $3.27\sigma$ & $1.59\sigma$ & $0.18\sigma$ & $0.49\sigma$ \\
        \hline \hline
    \end{tabular}
    }
    \caption{Predicted magnitudes of Hubble ``tensions'' with R21 for the cosmological models under consideration for source type No Delay and eLISA mission duration of 10 years, obtained from the three different methods.}
    \label{tab:tensions}
\end{table*}

In Table \ref{tab:tensions}, we summarize the status of the Hubble tension for the different models with respect to current datasets and each of the methods we have used, where we use the results for eLISA mission duration of 10 years and for the No Delay source type. Conclusions regarding the other cases can be easily obtained using the Gaussian tension (GT) metric
\begin{equation}
\text{GT}=\frac{\bar{x}_{i}-\bar{x}_{\text {SH0ES }}}{\left(\sigma_{i}^2+\sigma_{\text {SH0ES }}^2\right)^{1 / 2}}\:\:,
\end{equation}
where $\bar{x}_i$ and $\sigma_i$ are respectively the mean and standard deviation of observation $i$. We do not consider any other model selection technique in this work, a few of which are highlighted in \cite{H0_olympics}.

\subsection{Parametric methods}

We note that Fisher forecast indicates that future observations from eLISA would be able to constrain $H_0$ to much higher precision than current probes. In such a case, if the mean values of $H_0$ do not significantly shift from those obtained from the \textit{Planck} 2018 + BAO + Pantheon analysis, we expectedly see a rise in tension for each of the models. We also note that the error forecasts do not change significantly even if eLISA observes a different mean value (see Fig. \ref{fig:fishererrvsmean}). However, addressing tensions solely on the basis of inflated or deflated error bars is not the way to go. A shift in the mean value should ideally be taken into account, before one can comment on how well a given model helps alleviate the Hubble tension. It is unrealistic to assume that the mean of the $H_0$ posteriors would show no shift when subjected to actual eLISA data in the future. This is the primary limitation of the well-accepted and widely-used Fisher forecasting technique in the context of $H_0$ tension, although it still gives the best-case error constraint on cosmological parameters for a future mission, given its instrumental specifications and proper priors. 

Alternatively, one may use the conventional Markov Chain Monte Carlo (MCMC) technique for constraining cosmological parameters. This provides us with a handle on the mean values in addition to the errors. In the absence of real data, we apply MCMC to our simulated catalogs. In general, we note a slight decrease in tensions with R21 for our MCMC results compared to current constraints (see Table \ref{tab:gw_mcmc}). However, since an MCMC analysis assumes a model to obtain constraints on it from the data, one ends up with varying constraints for any given model when subjected to catalogs generated assuming different fiducial models. For example, when an MCMC analysis for the PEDE model is performed on the mock catalogs generated using the CPL model, the results vary significantly from the case where the analysis is done with the VM model catalogs instead. Since the whole point of trying to resolve the Hubble tension is to find a better alternative to $\Lambda$CDM, one should not arbitrarily choose catalogs to run the MCMC on. In this work, we quote the MCMC constraints such that the model assumed \textit{a priori} for the run is the same as the model which was used to generate the catalogs. This eliminates any bias to the $\Lambda$CDM model. This is particularly an important feature of our MCMC analysis which is in sharp contrast with most cases in the literature where $\Lambda$CDM fiducial values were almost always imposed on any arbitrary model attempting to going beyond $\Lambda$CDM, naturally leading to a biased analysis. 

Moreover, while our MCMC analyses have been done with the simulated GW data in isolation, we have used the priors from current datasets as inputs during catalog generation. Hence, we do not expect much deviation from our results if joint constraints are obtained using latest datasets and our simulated GW catalogs. The results thus obtained can be believed to be arising from a more or less robust analysis.

As for the status of the models themselves, we summarize a few salient features in the context of their tension-resolving potential. We discuss the results corresponding to the No Delay source type only as it provides a realistic middle ground among all the MBHB populations \cite{Ferreira}, for the conservative case of a 10-year mission duration. We notice the following trends with respect to existing \textit{Planck} 2018 + BAO + Pantheon constraints: 
\begin{enumerate}[(1)]
    \item Fisher forecasting shows higher tension, to varying degrees, for all the model parametrizations compared to current constraints. This is due to tighter error bounds. The tensions in case of $\Lambda$CDM, CPL, and JBP significantly worsen, but PEDE, VM, and VM-VEV do not show noticeable changes as their \textit{a priori} fiducial $H_0$ are closer to R21.
    \item MCMC results tend to increase the mean values by $\sim0.5$ km s\textsuperscript{-1} Mpc\textsuperscript{-1} for $\Lambda$CDM, CPL, JBP and PEDE, while for VM and VM-VEV they show a slight decrease. We observe an overall lowering of tension except in the case of VM-VEV.
\end{enumerate}

\subsection{Non-parametric method}

Non-parametric methods like GP have the ability to directly reconstruct the Hubble parameter $H(z)$ from the mock catalogs. The intercept at $z=0$ hence gives the value of the Hubble constant. The results are then solely dependent on the simulated data used to ``train'' the GP algorithm. We employ this method in Sec. \ref{sec:GP} to constrain the posteriors on $H_0$ by training the machine on 500 generated catalogs for every choice of fiducial (corresponding to a specific model from Sec. \ref{sec:models}), MBHB source type and eLISA mission duration. In general, we find that GP makes the constraints on the parameters for each model relatively wider than what was obtained from current datasets. The uncertainties are also wider than what we obtained using the Fisher forecasting method and from the MCMC runs. But unlike MCMC, GP can quote mean values and errors in a non-parametric way. This is the primary advantage of using GP over the other two. Moreover, GP also tends to shift the mean values slightly (see Table \ref{tab:gp_H0}). The significance of these mean shifting tendencies merits further investigation in the light of other missions as well as via comparative studies incorporating other ML algorithms, like neural networks or genetic algorithms \cite{ML_Cora}.

For the conservative case of a 10-year mission duration and focusing once again on the No Delay source type, we notice that for the fiducials F2 and F3, the mean value of $H_0$ reconstructed with GP tends to be higher by $\sim1$ km s\textsuperscript{-1} Mpc\textsuperscript{-1}. Using the F4 fiducial, we find that shifting of the mean $H_0$ is in the opposite direction but is rather insignificant. As for the fiducials F5 and F6, we observe a similar lowering of the mean values by about $\sim0.5$ km s\textsuperscript{-1} Mpc\textsuperscript{-1}.

\subsection{Comparison and challenges}

Based on Table \ref{tab:tensions}, we now proceed to elucidate the behavior of each model and its corresponding fiducial across the three different methods:
\begin{enumerate}[(1)]
    \item With F1 ($\Lambda$CDM), F2 (CPL) and F3 (PEDE), tensions decrease progressively from Fisher to MCMC to GP. For GP, the alleviation is primarily because of the shifting of the mean to somewhat higher values. For MCMC, this is in some part due to higher mean values, and partly due to inflated errors.
    \item Although MCMC shifts the mean for F4 (JBP) to a slightly higher value than GP, the smaller error bar associated with the former compensates for the shift. This renders the tensions predicted by MCMC and GP roughly equal.
    \item For F5 (VM) the tensions also decrease progressively from Fisher to MCMC to GP, due to a tendency to slightly lower the mean $H_0$ compared to that from current datasets.
    \item With F6 (VM-VEV), the tensions are higher for GP and MCMC than for the Fisher forecast and current constraints. This is because GP shows a tendency to slightly lower the mean $H_0$, which from current datasets is very close to R21.
\end{enumerate}

\begin{figure*}[!htb] 
\begin{center} 
\begin{subfigure}{.3\textwidth}
        \centering
        \includegraphics[width=\textwidth]{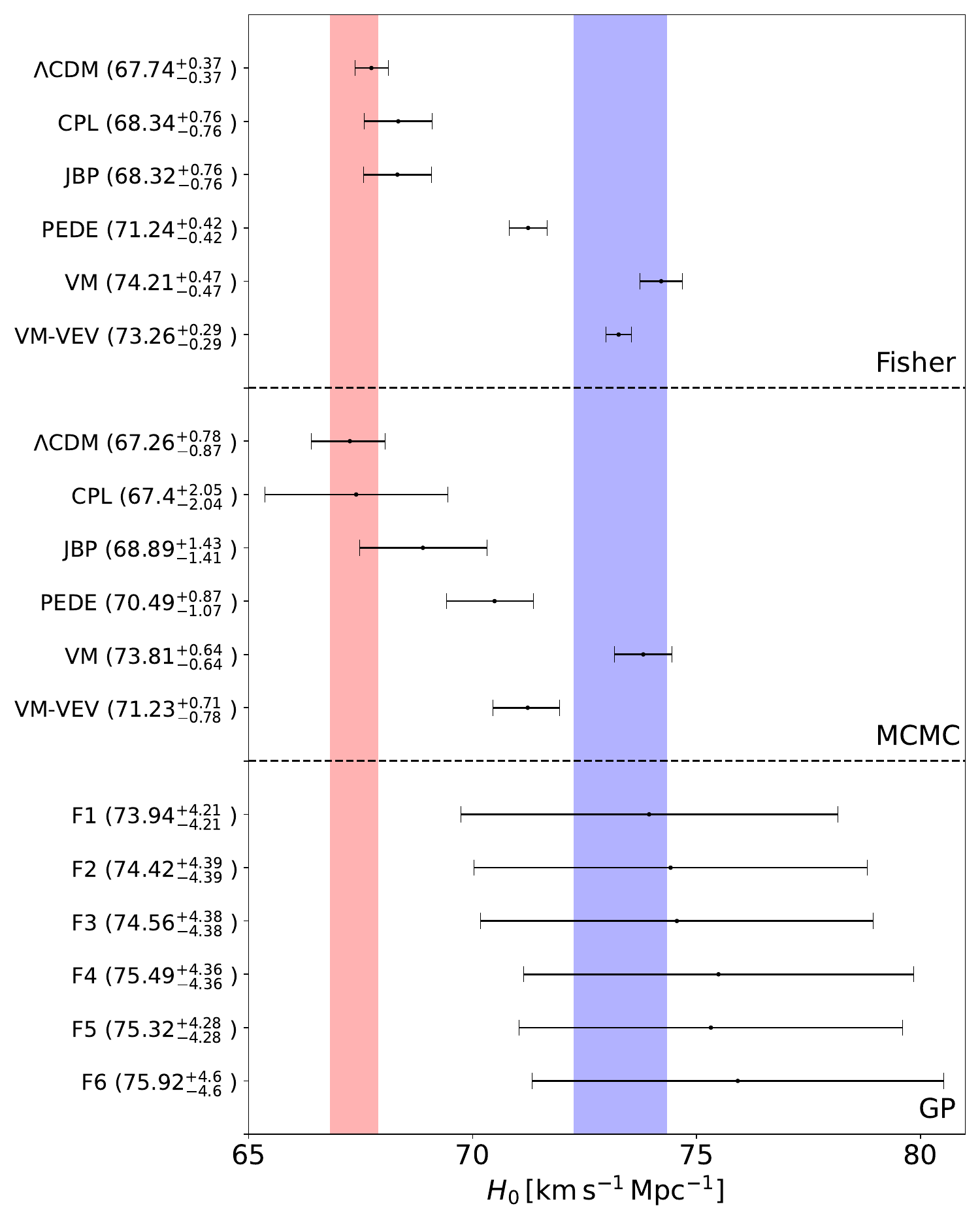}
        \caption{5 Years Delay}
    \end{subfigure}
    \begin{subfigure}{.3\textwidth}
        \centering
        \includegraphics[width=\textwidth]{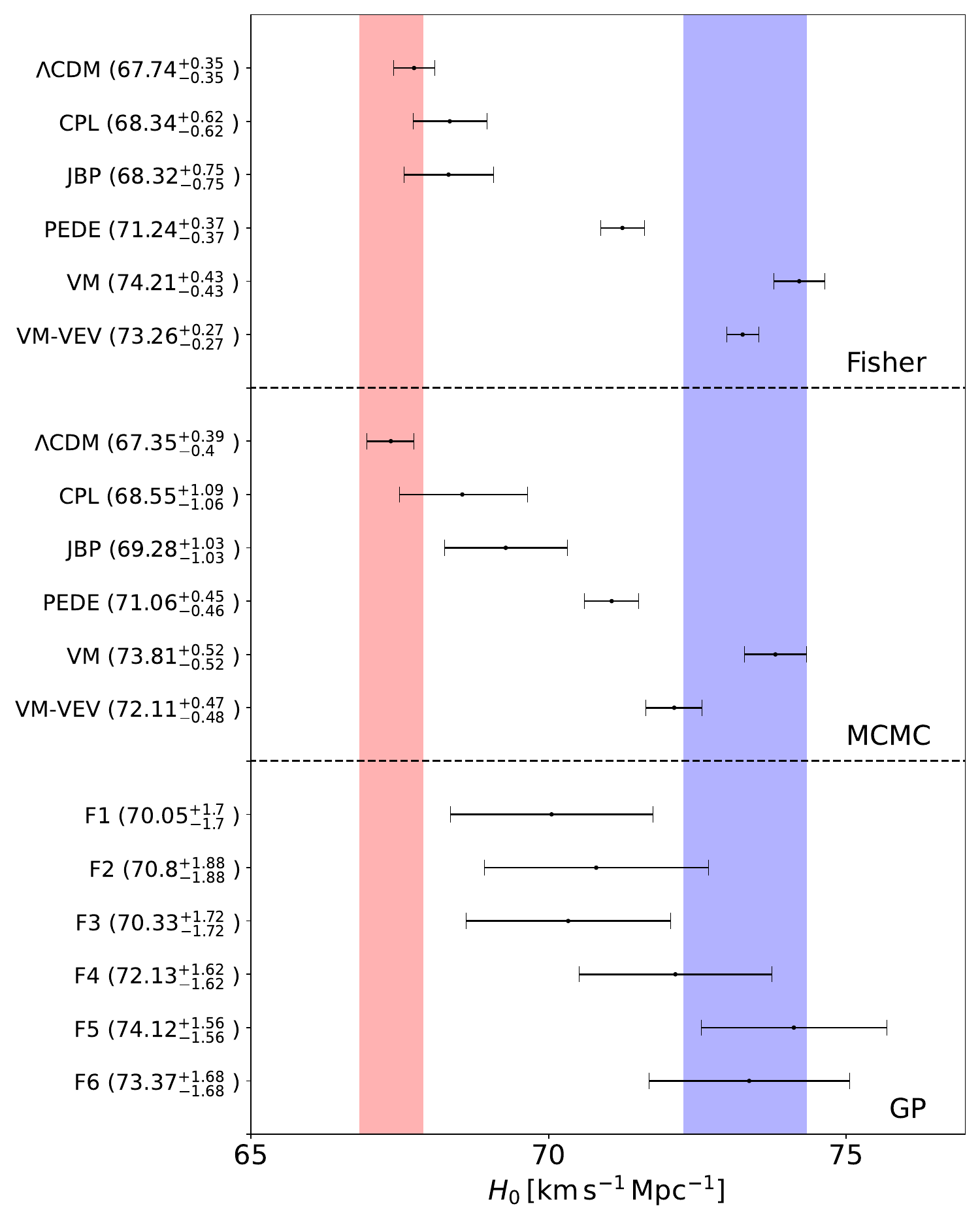}
        \caption{10 Years Delay}
    \end{subfigure}
    \begin{subfigure}{.3\textwidth}
        \centering
        \includegraphics[width=\textwidth]{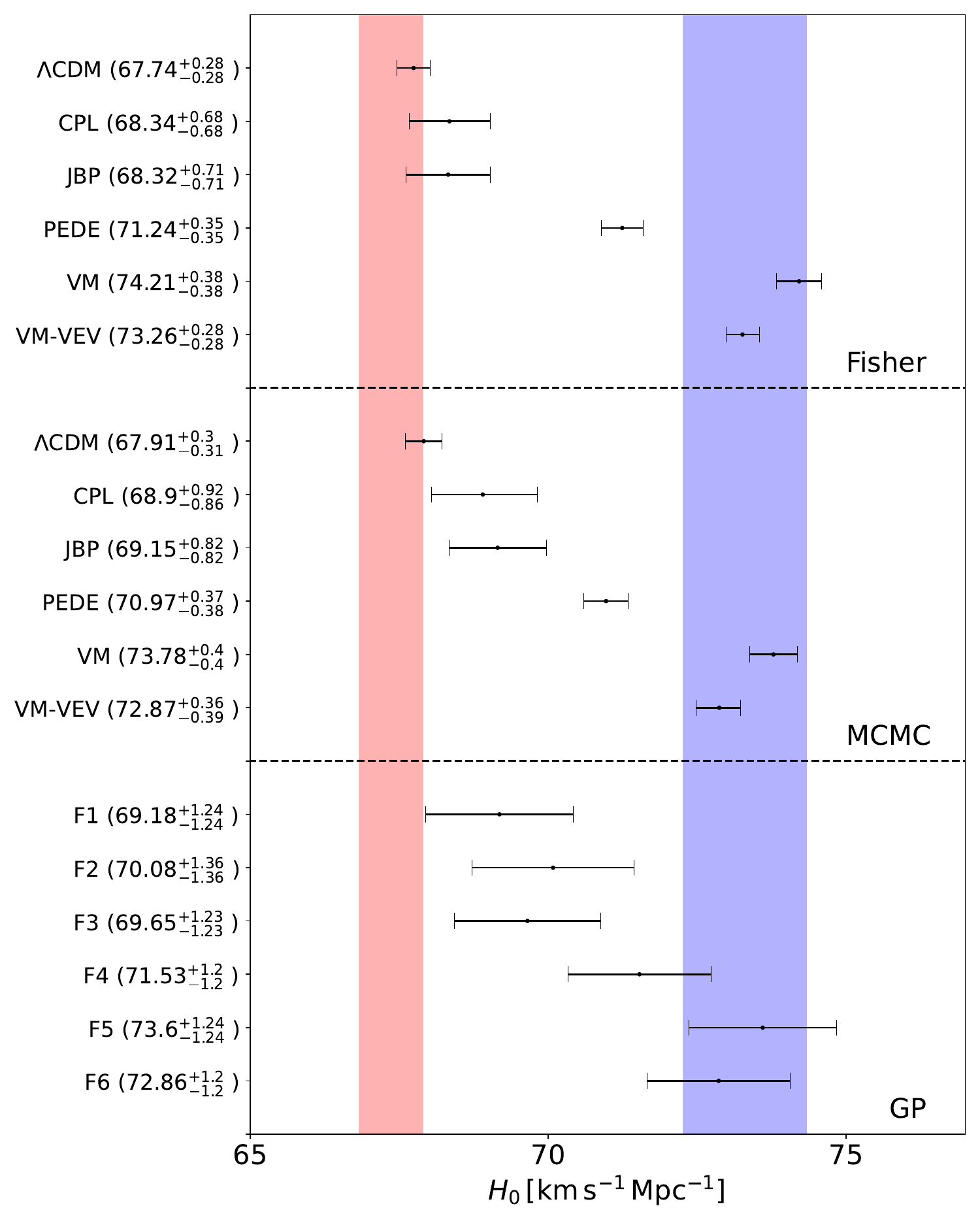}
        \caption{15 Years Delay}
    \end{subfigure}
    \begin{subfigure}{.3\textwidth}
        \centering
        \includegraphics[width=\textwidth]{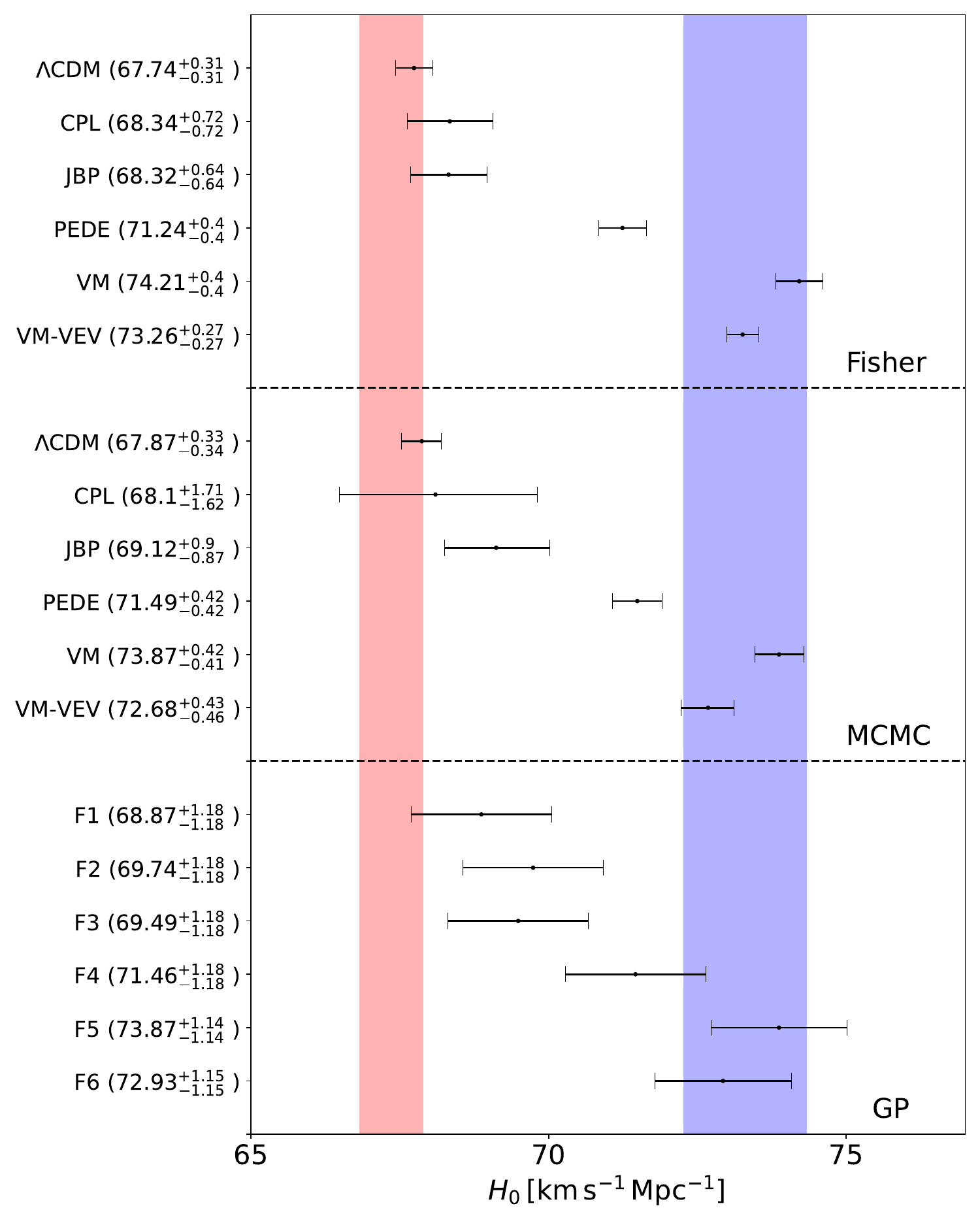}
        \caption{5 Years No Delay}
    \end{subfigure}
    \begin{subfigure}{.3\textwidth}
        \centering
        \includegraphics[width=\textwidth]{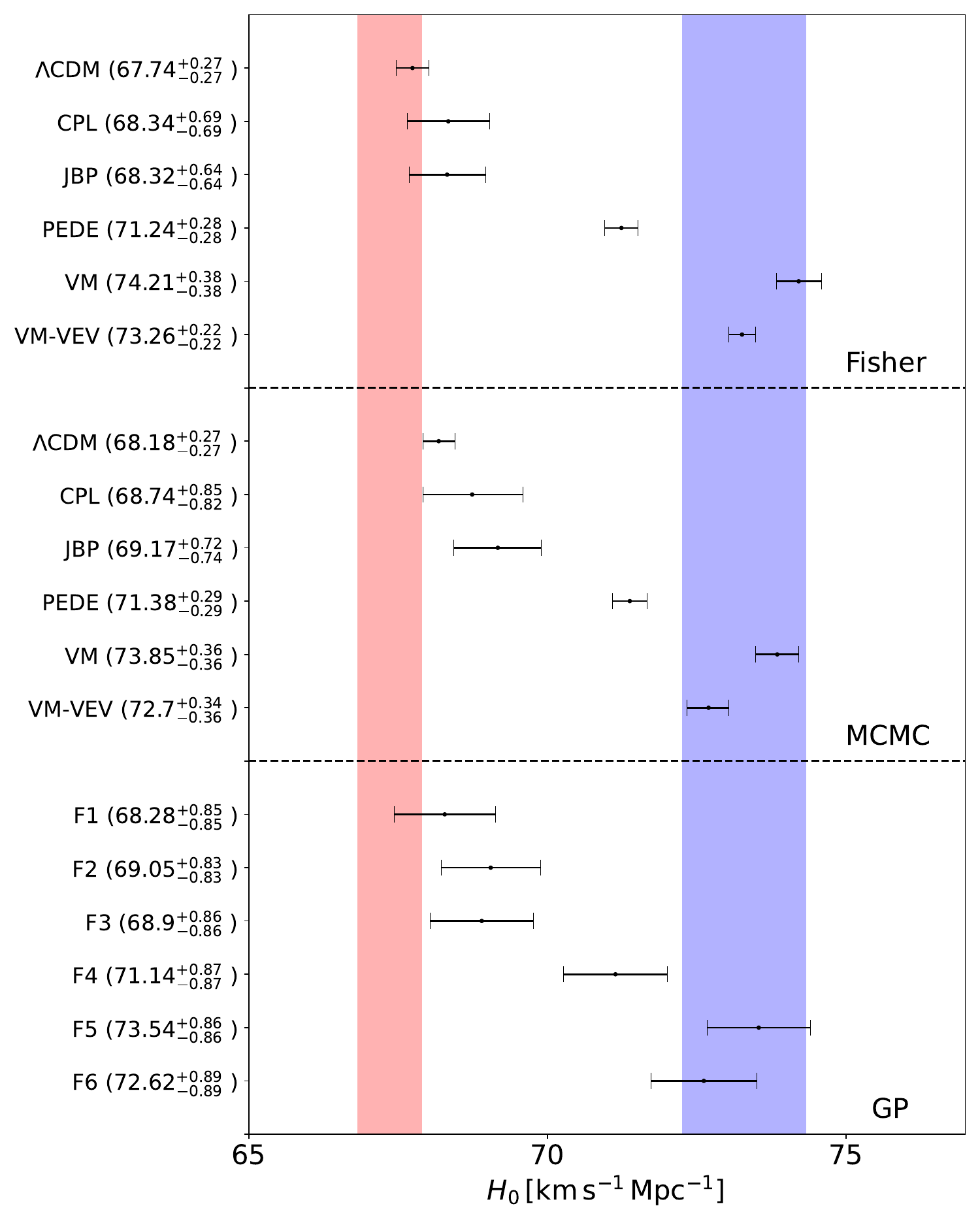}
        \caption{10 Years No Delay}
    \end{subfigure}
    \begin{subfigure}{.3\textwidth}
        \centering
        \includegraphics[width=\textwidth]{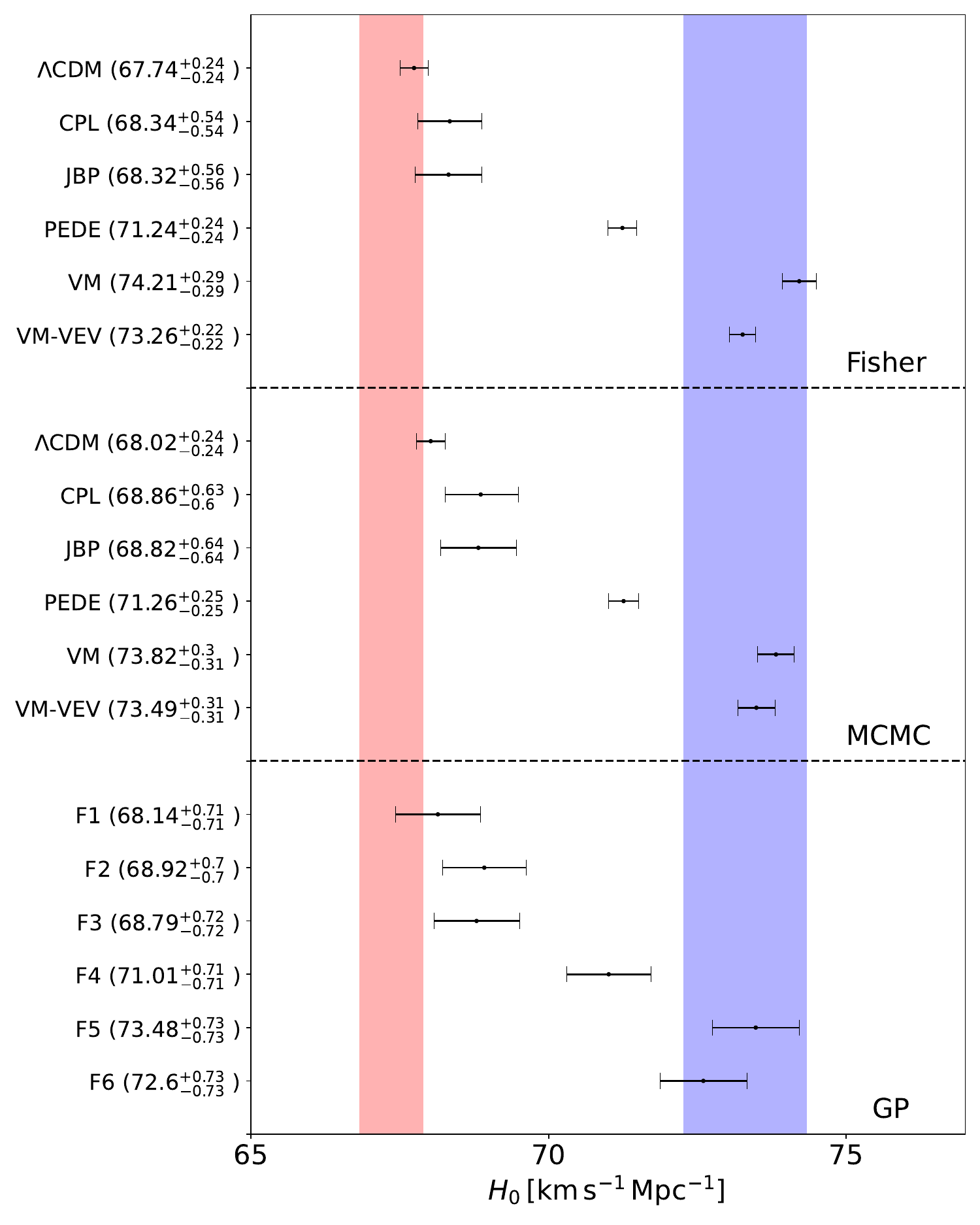}
        \caption{15 Years No Delay}
    \end{subfigure}
    \begin{subfigure}{.3\textwidth}
        \centering
        \includegraphics[width=\textwidth]{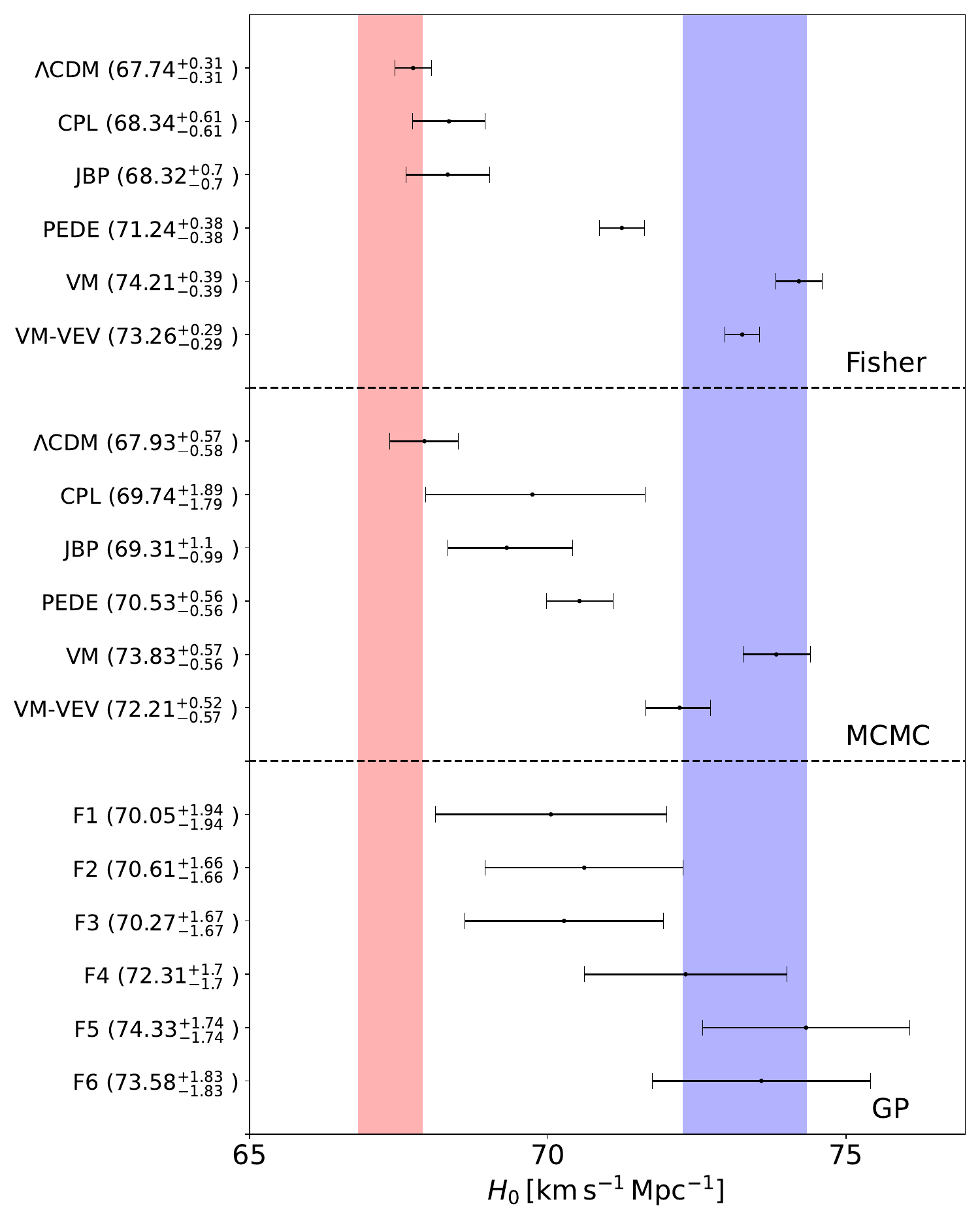}
        \caption{5 Years Pop III}
    \end{subfigure}
    \begin{subfigure}{.3\textwidth}
        \centering
        \includegraphics[width=\textwidth]{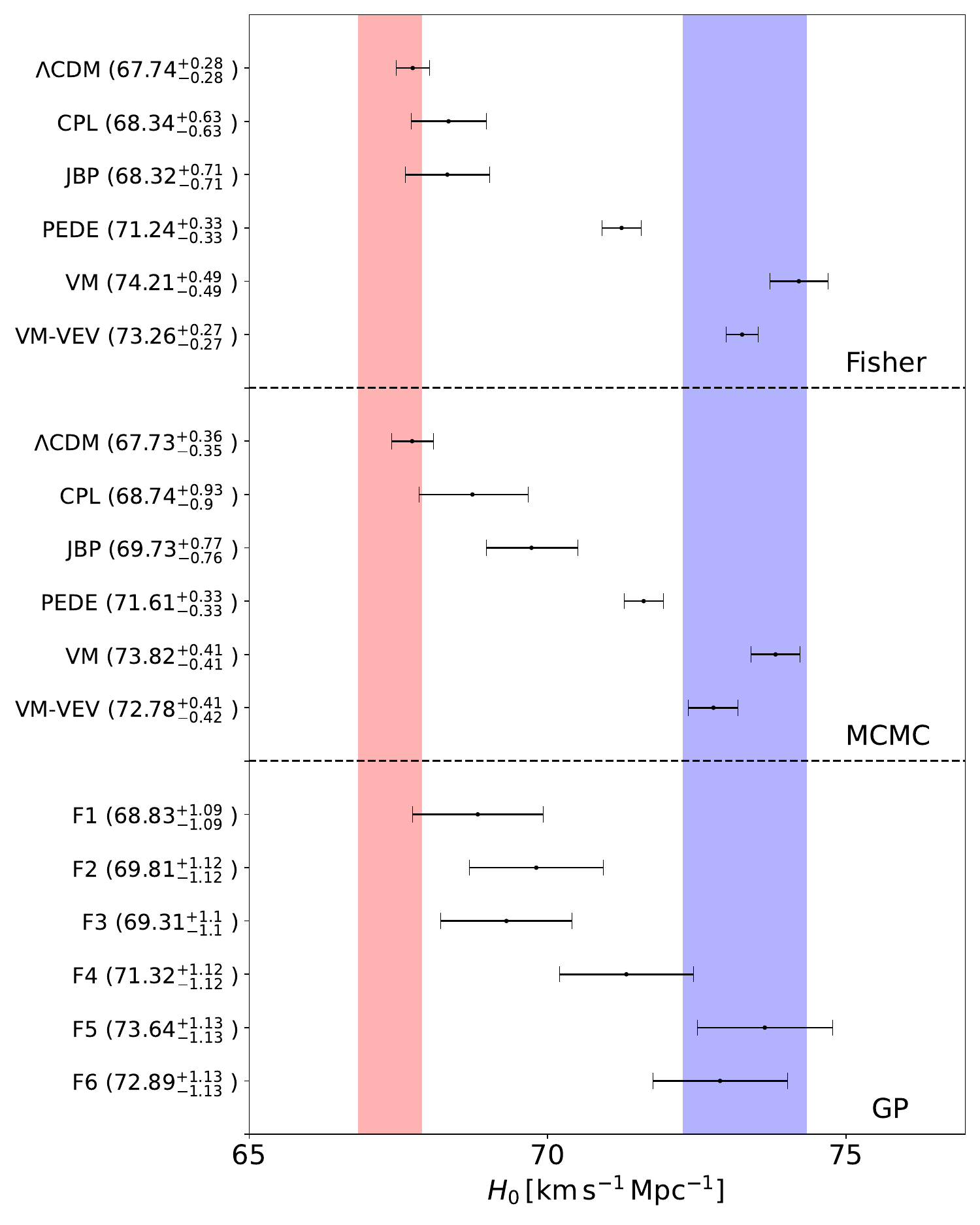}
        \caption{10 Years Pop III}
    \end{subfigure}
    \begin{subfigure}{.3\textwidth}
        \centering
        \includegraphics[width=\textwidth]{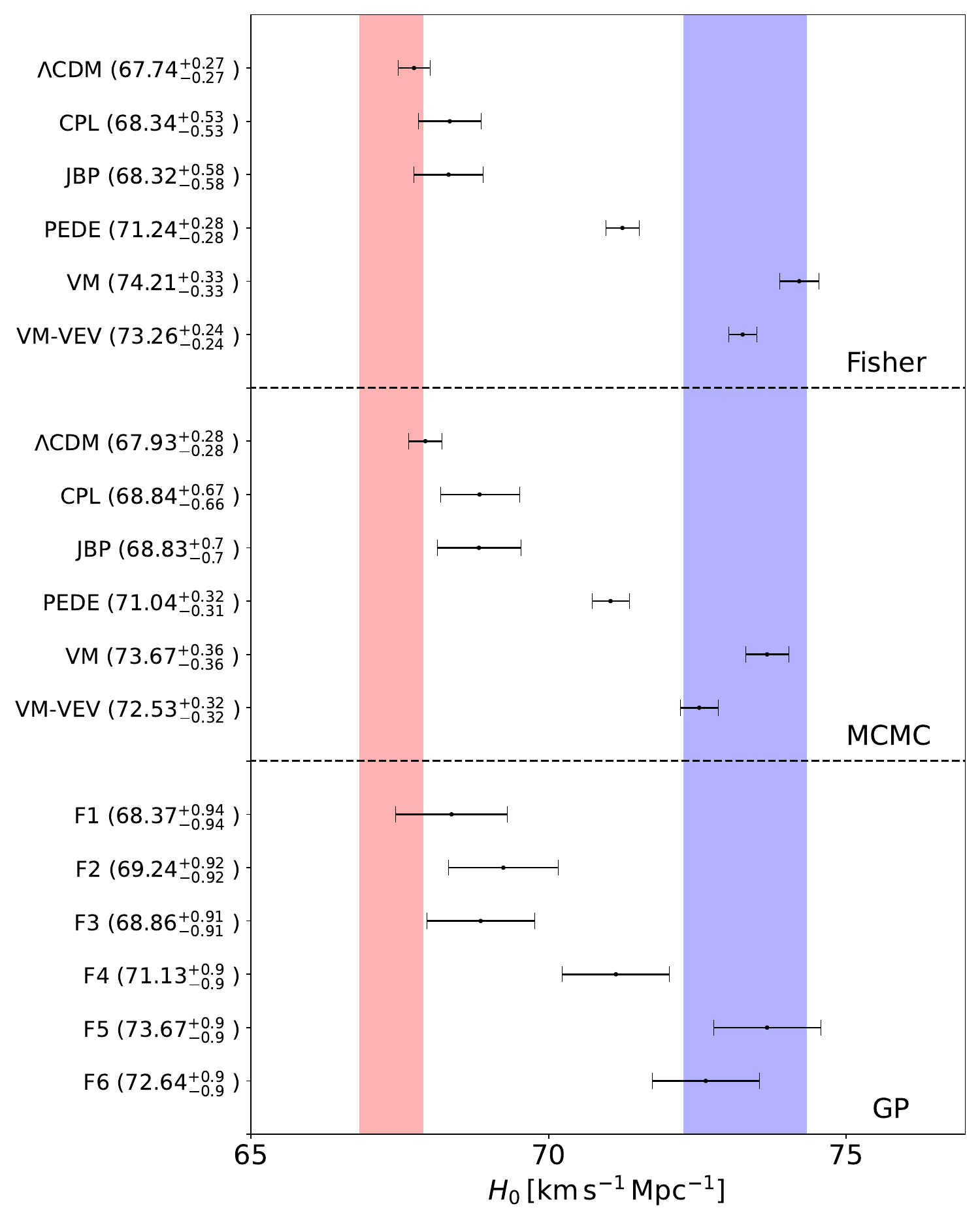}
        \caption{15 Years Pop III}
    \end{subfigure}
    \caption{Whisker plots of eLISA forecasts on $H_0$ (in units of km s\textsuperscript{-1} Mpc\textsuperscript{-1}) by the three different methods. The set of fiducials F1, F2, F3, F4, F5, F6 used for GP has been defined in Sec. \ref{sec:GP}.}
    \label{fig:whisker}
\end{center}
\end{figure*}

Figure \ref{fig:whisker} depicts the behaviors of the different models for all three source types and mission durations, with the blue bar representing the latest SH0ES constraints, and the red bar, the Planck 2018 constraints. This is basically an extension of the above-mentioned points (based on Table \ref{tab:tensions} only), keeping the major conclusions unaltered. However, we reiterate that one should not view the performance of any cosmological model in the context of $H_0$ tension in isolation. For example, the PEDE, VM and VM-VEV models show highly reduced tensions with R21 for all cases, but they cannot necessarily be concluded as better alternatives to $\Lambda$CDM. The PEDE model shows promise of resolving tension at the background level but falters when fitting current clustering data where it is not as efficient as $\Lambda$CDM \cite{Rezaei2020}. The VM and VM-VEV models, on the other hand, have been shown to suffer from a poorer goodness of fit to current data compared to $\Lambda$CDM \cite{VM_Di_Valentino_2020}, as highlighted earlier in Sec. \ref{sec:models}. So, for any particular model or a class of models to pass the real test of the $H_0$ tension (or any other tension as such), an in-depth study considering all the pros and cons of the model \textit{vis-\`{a}-vis} other models is necessary. 

Moving on to a comparison among the methods themselves, we first focus on the two parametric approaches. The Fisher approach is a computationally quick parametric tool which is relatively independent of the catalogs and reliant more on the errors induced by the instrumental specifications for a mission, and requires as input an assumed model of cosmology. Thus, Fisher analysis is useful for sensitivity forecasts if the likelihood for the parameter space is known up to good approximation. If we have a good prior handle on the mean, Fisher forecasting is by far the best predictive tool for error estimation. In contrast, the MCMC results simultaneously give us a handle on the mean in addition to the errors, depending on the model assumed. Moreover, although the inverse Fisher matrix gives some approximate one-dimensional confidence regions and two-dimensional elliptic contours, Fisher analysis essentially relies on the Gaussian approximation for the posterior distribution. So, if one is interested in the details of the posterior such as the correlations between model parameters with some skewness or kurtosis, then Bayesian inference via MCMC needs to be employed. We note that a few of the posterior distributions obtained through MCMC in Sec. \ref{sec:GWMCMC} show slight deviations from Gaussianity. While this should, in principle, entail higher-order corrections to the standard Fisher formula, the imprints of such corrections are expected to be sub-dominant in the current scenario. This is further evident as the $1\sigma$ error forecasts from Fisher analysis (in Table \ref{Tab:Fisher}) are close to those obtained from MCMC (in Table \ref{tab:gw_mcmc}). To gain more insight into the validity of cosmological Fisher matrix forecasts and its comparison to MCMC, one can refer to \cite{Wolz:2012sr, Bose:2019psj, Yahia-Cherif:2020knp} and references therein.

On the other hand, we see that the advantage of GP lies in getting an estimate on both the mean and the error in a \textit{non-parametric} manner. Our results show the behavior of the reconstructed Hubble parameter across a range of well-motivated fiducials, instead of assuming any particular fiducial {\em a priori} that might lead to biased results. GP tends to give wider error bars compared to Fisher when subjected to the same amount of data because there is no input data for ${d_L}'$ while training. Furthermore, by construction, GP is not too confident with reconstructing $H(z)$ where data is relatively sparse such as at higher redshifts, which is evident at $z>4$ in Fig. \ref{fig:gp_Hz}. A GP reconstruction is computationally quicker if the training is done using an optimized approach. Finally, the major disadvantage of GP is its kernel dependence \cite{OColgain:2021pyh}. In order to get robust results, a marginalization over the hyperparameters is the way through, although this makes it quite computationally expensive. Having said that, we must appreciate that among the three approaches, GP (or any ML tool as such) has the least bias from particular models and hence this direct reconstruction technique merits further exploration.

We also notice generically for all the methods that a longer mission duration results in tighter constraints, as do source types that produce a larger number of detectable events. This is the expected trend in any observational scenario. In the case of simulated data and machine learning techniques, one needs to be cautious so as not to confuse higher precision with overfitting. We acknowledge this as an open issue which warrants a better understanding of the caveats of various ML algorithms when applied to cosmology. Also, we consider a finite number of catalogs in each of the three branches of our method, which might introduce some bias to our results. Repeating the whole exercise with a much larger number of catalogs may help increase confidence in our conclusions. But we do not expect them to provide tighter constraints on the uncertainties.

In a nutshell, our overall impression on the three different approaches is as follows. Although the majority of the models considered in this work show somewhat relaxed tensions with R21 in general, we cannot stress on it strongly because of the relatively wider error bounds than those from current datasets. However, we suggest that any comment regarding the Hubble tension for a future mission should not be made solely on the basis of Fisher analysis, and robust Bayesian methods like MCMC be used to effectively make conclusions by taking into account both the mean values and the errors. Additionally, GP also provides us with estimates for both the mean values and the associated uncertainties via a reconstruction of the evolutionary history directly from data, without bias towards any particular model. In this manner, a comparative study involving different methods will eventually lead to a robust and more scientific analysis of any particular model or any particular mission.


\section{\label{sec:conclusion}Conclusion and future directions}

The presently observed Hubble tension might be quite generic to current datasets \cite{Bhattacharyya_2019}. This prompts us to look at future missions. In this paper, we have focused on the prospects of future GW observations in alleviating this tension, taking eLISA as a specific example on a wide class of cosmological models/parametrizations, namely, $\Lambda$CDM, CPL, JBP, PEDE, VM, and VM-VEV. We perform our three-pronged approach, namely, (i) Fisher forecast, (ii) Markov Chain Monte Carlo and (iii) Gaussian Processes in machine learning, and our key findings include the following:
\begin{enumerate}[(1)]
    \item We bring our chosen models to an equal footing using Markov Chain Monte Carlo analysis based on \textit{Planck} 2018 + BAO + Pantheon. In particular, we emphasize that the PEDE model cannot resolve tension with the SH0ES measurement within $1\sigma$ (as claimed earlier in \cite{YANG2021100762}) without the R21 prior. Also, the JBP parametrization is not as efficient in alleviating the Hubble tension with R21, as was previously concluded in \cite{Yang2021} without considering SNIa data in the analysis. We have also used these latest constraints as fiducial values in subsequent analysis in order to arrive at up-to-date conclusions.
    \item We have then performed a conventional forecasting on eLISA by taking into account all the models under consideration using the Fisher matrix method, which evidently shows that it will be able to constrain the Hubble constant to a much higher precision than the \textit{status quo}. From our study, we infer that eLISA would serve as a powerful mission when it comes to constraining the Hubble constant to well below the percent level.
    \item Constraining our chosen models with MCMC using only the respective simulated catalogs show slightly reduced tensions for most of the models. This is attributed to MCMC's ability to provide a handle on the mean values in addition to the errors.
    \item Our preliminary analysis using GP shows a slight trend for the $H_0$ values to tend towards the locally measured values, in spite of us not including any prior from direct measurements in our analysis. These trends merit further investigation. Among the three approaches, GP (or any ML tool as such) has the least bias from particular models, with the model-dependent input in our scheme of work being the fiducials used to generate the synthetic GW catalogs to run GP on. Thereafter, there is no model-dependence in the reconstruction pipeline. Hence, this direct reconstruction technique needs to be explored further.
    \item  Combining all the results and analysis therefrom, we come to the conclusion that any comment regarding the Hubble tension for a future mission should be carried out by employing all three methods so as to make the analysis scientifically more appealing, until the community is sure about a strong and competitive advantage of a particular method over the others.
\end{enumerate}
 
Of course, there is considerable scope for future works, both in terms of improvements in precision and in terms of exploring other avenues. Although we have prepared the catalogs in as realistic a way as possible, the method of catalog generation might be further refined by carefully taking into account detailed astrophysics of possible GW sources and various instrumental sensitivities. This essentially needs a collaboration between the cosmology and numerical relativity communities, which we hope, would take place sometime in the near future. To arrive at more realistic forecasts, one should also, in principle, marginalize over properties of the MBHB population models that cannot be measured directly in an observational setup, \textit{e.g.} the delay time which distinguishes the Delay and No Delay types. Such marginalization might predict somewhat greater uncertainty in the final results, resulting in wider confidence intervals and/or decreased signal-to-noise ratios for the detected signals. The results may also depend on the particular marginalizing technique to be used. For example, certain grid-based or Monte Carlo-based methods may entail additional computational complexity and may even require extra assumptions about the source type distributions. In this work, we have not explored the effects of such marginalization, as we have opted for a case-by-case analysis of the three distinct MBHB source types and the constraining power of their catalogs. We plan to focus on the prospects and consequences of marginalization over source properties in a future study.

Secondly, in this work we only look into the prospects of eLISA, and in particular, with the L6A2M5N2 configuration. A comparative study can and must be done for other configurations using our proposed methodology. Moreover, in the present study our focus was on intermediate redshifts, and eLISA is the next upcoming mission in this direction.  There are multiple other planned next generation GW missions, \textit{e.g.}, DECIGO, ET and the Big Bang Observer (BBO) \cite{bbo1,bbo2} among others, which plan to probe different redshifts and hopefully more events. Our analysis can be extended to those missions as well. 

Further, we have assumed all the observable events to be accompanied by EM emissions, \textit{i.e.}, bright sirens. If in the future galaxy correlation methods improve, dark sirens would provide us with even more GW observations and help constrain the Hubble constant in a stronger manner. This would especially be useful for ML algorithms, the results of which steadily improve as the quantity of training data is increased.

GP is not the only applicable ML tool at hand. There exists a variety of other ML algorithms in the literature, many of which have already found applications in various fields of cosmology \cite{ML_Cora}. Some of these may be parametric. Others, like GP, may be non-parametric - such as neural networks and its derivatives. However, they require much larger amounts of data to train and are more computationally expensive than GP in general. Nevertheless, they seem to be extremely promising for future cosmological studies, some of which we plan to explore in future works.

On the theoretical side, GW luminosity distance $d_L^{(GW)}$ can differ from $d_L$ in alternative theories of gravity. This is why standard sirens can also serve as a powerful observational probe of modified gravity \cite{modgrav1,modgrav2,modgrav3}. Thus, one can extend the present work by moving beyond GR and employing our current method of study to forecast on aspects of modified gravity models as well.

Last but not the least, precision cosmology in the present era does not only suffer from the Hubble tension but also from numerous other issues \cite{Abdalla_2022}. These tensions are often degenerate and the resolution of one can worsen the other, as has been seen in most cases. While we have looked at the Hubble tension in isolation in this work, a more complete analysis by taking into account other related tensions using the current methodology, must be carried out in order to comment on the models more effectively. For example, the current study may be extended by considering large scale structure (LSS) missions such as Euclid \cite{euclid}, both present and future, to constrain and comment on the $S_8$ tension simultaneously with that of $H_0$.

Finally, the current analysis can be generalized beyond GW observations alone. Fortunately, at intermediate redshifts, GWs are not the only future probe. The Epoch of Reionization (EoR), in particular, would be probed by future radio interferometric observatories such as the Square Kilometre Array (SKA) \cite{ska}. This would serve as a complementary tool to GWs for this era, and will help constrain the Hubble constant at intermediate redshifts much more consistently, the prospects of which we leave for future work.


\section*{Appendix: Effects of arbitrary catalog fiducials on forecast results}

\begin{figure*}[!ht]
    \centering
    \includegraphics[width = 0.485\textwidth]{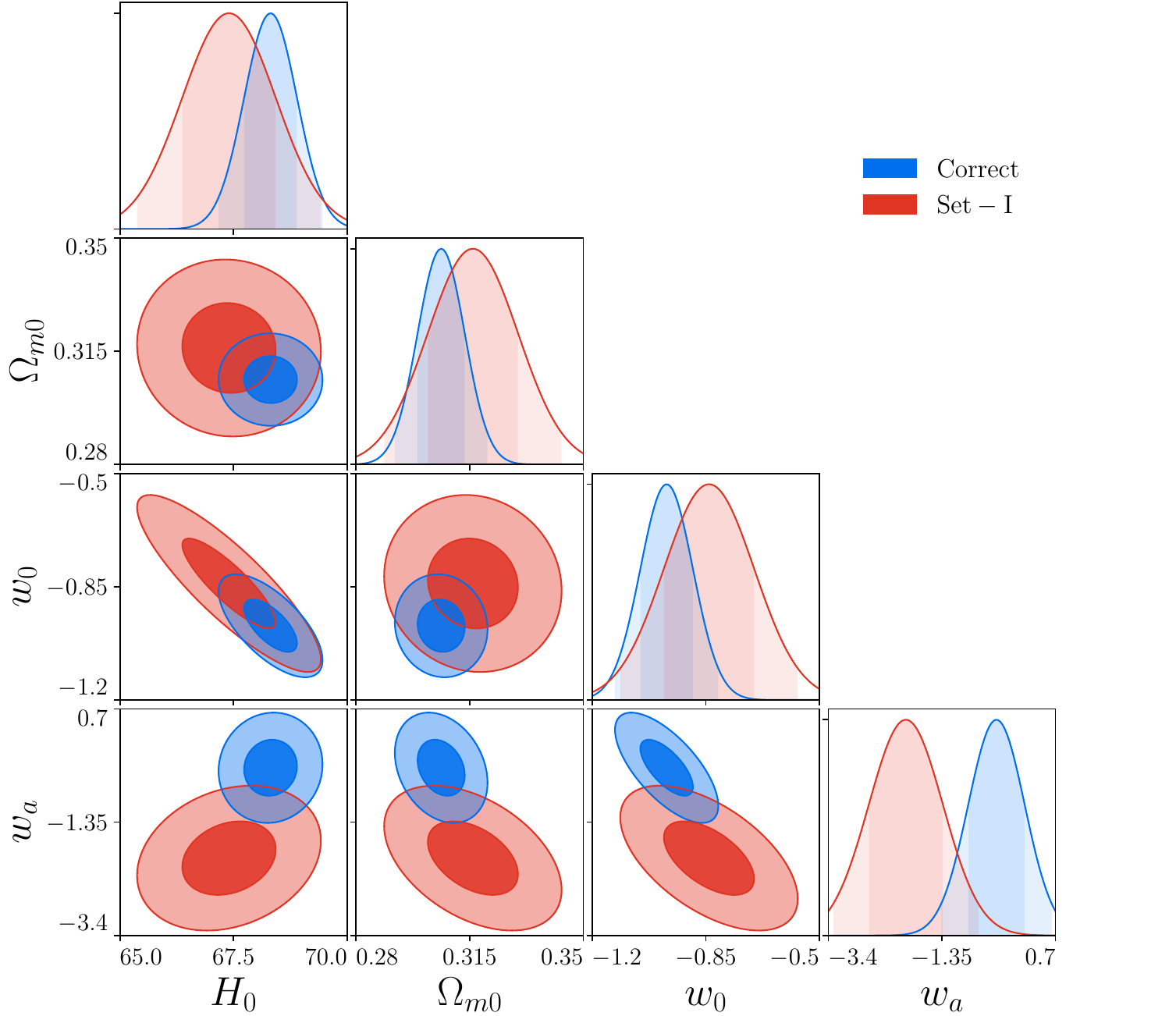}
    \includegraphics[width = 0.485\textwidth]{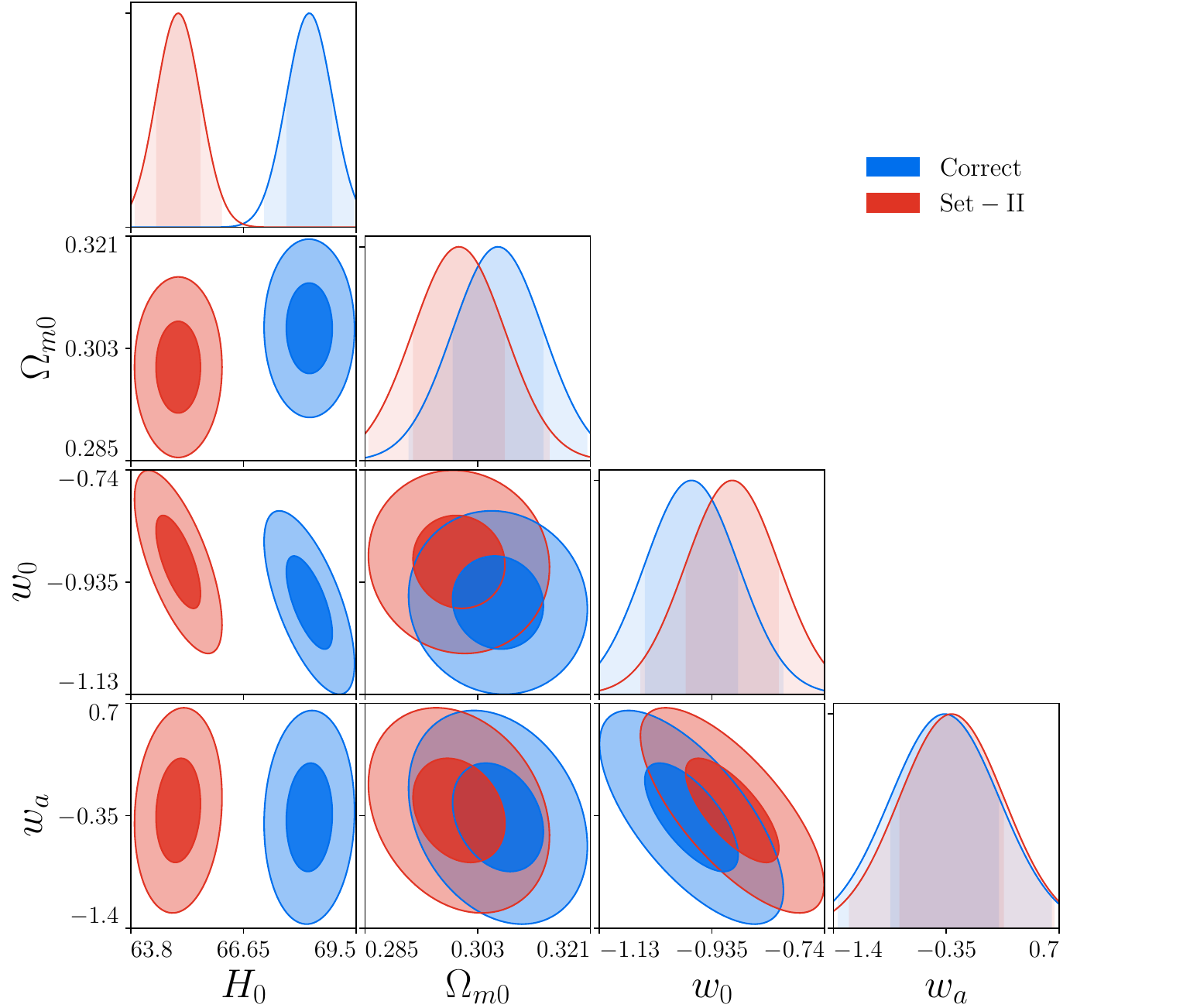}
    \caption{Plot showing the sensitivity of Fisher analysis to both the mean and the 1$\sigma$ error fiducials (represented by Set-I from Table \ref{tab:wrong_fids}) in the left panel, and the effect of incorrectly chosen fiducial mean values (represented by Set-II from Table \ref{tab:wrong_fids}) on Fisher analysis results in the right panel. Both plots are for the JBP model on considering 10-year catalogs of the No Delay source type.}
    \label{fig:wrong_plot_fisher}
\end{figure*}

In this appendix, we demonstrate the effect of catalogs, constructed on the basis of incorrectly chosen fiducials, on the forecast results. This exercise further establishes the necessity of constraining cosmological models on the basis of complete and consistent datasets as done in Sec. \ref{sec:latestdata}, before those constraints are used as fiducials in forecasting studies. For the purpose of illustration, we take the JBP model as an example here, that was constrained earlier in \cite{Yang2021} on the basis of \textit{Planck} 2018 + BAO.

We construct two different 10-year mock catalogs corresponding to the No Delay source type, the first one based on the inadequate constraints from \cite{Yang2021} which we refer to as Set-I, and the second one on some fictitious and arbitrarily chosen fiducials which we refer to as Set-II. The values of these fiducials, alongside the correct ones for JBP as obtained in Sec. \ref{sec:latestdata}, are listed in Table \ref{tab:wrong_fids}. For Set-II, the $1\sigma$ errors have been kept identical to those of the correct constraints in order to clearly illustrate the effect of mean shifts on the forecast results, while keeping the level of precision unchanged from the actual constraints. 

\begin{table*}[!ht] 
   \resizebox{\textwidth}{!}{\renewcommand{\arraystretch}{1.6} \setlength{\tabcolsep}{20 pt} \centering
    \begin{tabular}{c | c c c c }
        \hline\hline
        \textbf{Fiducials} & $H_0$  [km Mpc$^{-1}$ s$^{-1}$] & $\Omega_{m0}$ & $w_0$ & $w_a$ \\ \hline	
        Set-I & $67.40\pm2.40$ & $0.316\pm0.023$ & $-0.84\pm0.30$ & $-2.00\pm0.75$ \\ \cline{1-5}
        Set-II & $65.00\pm0.81$ & $0.300\pm0.008$ & $-0.90\pm0.12$ & $-0.30\pm0.75$ \\ \cline{1-5}
        Correct & $68.32\pm0.81$ & $0.306\pm0.008$ & $-0.97\pm0.12$ & $-0.36\pm0.75$ \\
        \hline \hline
    \end{tabular}
    }
    \caption{Two sets of improper fiducial parameter values chosen to demonstrate the effect of fiducials on the catalogs and subsequent forecast results: Set-I is borrowed from \cite{Yang2021} and Set-II constitutes some arbitrarily chosen values. The last row lists the JBP constraints from Sec. \ref{sec:latestdata} for ready reference.}
    \label{tab:wrong_fids}
\end{table*}

Figure \ref{fig:wrong_plot_fisher} shows side-by-side comparisons between the Fisher results corresponding to the correct set of fiducials (from Fig. \ref{fig:fishernodelay}) and the two sets from Table \ref{tab:wrong_fids}. This demonstrates the sensitivity of forecast results to the choice of fiducial values used to construct the mock GW catalogs. The differences are at the level of the projected constraining capabilities of eLISA. Thus, an improper choice of catalog fiducials is likely to result in an erroneous forecast analysis.
\\
\\
\noindent\textbf{Data/Code availability}: The codes and data may be made available upon reasonable request.


\acknowledgments

We gratefully acknowledge the use of the publicly available codes \href{https://github.com/lesgourg/class_public}{\textit{CLASS}}, \href{https://github.com/brinckmann/montepython_public}{\textit{MontePython}}, \href{https://github.com/CosmicFish/CosmicFish}{\textit{CosmicFish}}, \href{https://github.com/dfm/emcee}{\textit{emcee}}, \href{https://github.com/cmbant/getdist}{\textit{GetDist}} and \href{https://github.com/carlosandrepaes/GaPP}{\textit{GaPP}}.
We thank Prasanta Dutta for preliminary discussions on ML. We also thank the anonymous reviewer for their valuable suggestions towards the improvement of the manuscript. RS thanks ISI Kolkata for financial support through Senior Research Fellowship. AB thanks CSIR for financial support through Junior Research Fellowship (File no. 09/0093(13641)/2022-EMR-I). PM thanks ISI Kolkata for financial support through Research Associateship. SP thanks the Department of Science and Technology, Govt. of India for partial support through Grant No. NMICPS/006/MD/2020-21. We acknowledge the computational facilities of ISI Kolkata.


\bibliographystyle{JHEP} 
\bibliography{biblio} 

\end{document}